\documentclass{aa}  

\usepackage{graphicx}
\usepackage{txfonts}
\usepackage{hyperref}
\begin{document}

   \title{Solar Orbiter Observations of the Kelvin-Helmholtz Instability in the Solar Wind}

   \author{R. Kieokaew\inst{1}, B. Lavraud \inst{1,2}, Y. Yang \inst{3}, W. H. Matthaeus \inst{4}, D. Ruffolo \inst{5}, J. E. Stawarz \inst{6}, S. Aizawa \inst{1}, C. Foullon\inst{7}, V. Génot\inst{1}, R. F. Pinto\inst{1,8}, N. Fargette \inst{1}, P. Louarn \inst{1}, A. Rouillard\inst{1}, A. Fedorov\inst{1}, E. Penou\inst{1}, C.J. Owen\inst{9}, T. Horbury\inst{6}, H. O’Brien\inst{6}, V. Evans\inst{6}, and V. Angelini\inst{6} 
          }

   \institute{\inst{1}Institut de Recherche en Astrophysique et Planétologie, CNRS, UPS, CNES,
              9 Ave. du Colonel Roche 31028 Toulouse, France \email{rkieokaew@irap.omp.eu} \\
         \inst{2}Laboratoire d'Astrophysique de Bordeaux, Univ. Bordeaux, CNRS, B18N, allee Georoy Saint-Hilaire, 33615 Pessac, France \\
         \inst{3}Department of Mechanics and Aerospace Engineering, Southern University of Science and Technology, Shenzhen 518055, People’s Republic of China \\
         \inst{4}Department of Physics and Astronomy and Bartol Research Institute, University of Delaware, Newark, DE 19716, USA \\
         \inst{5}Department of Physics, Faculty of Science, Mahidol University, Bangkok 10400, Thailand \\
        \inst{6}Space and Atmospheric Physics, Department of Physics, Blackett Laboratory, Imperial College London, London, UK SW7 2AZ \\
         \inst{7}CGAFD, Mathematics, CEMPS, University of Exeter, Exeter, UK \\
        \inst{8}LDE3, DAp/AIM, CEA Saclay, 91191 Gif-sur-Yvette, France \\   
        \inst{9}Department of Space and Climate Physics, University College London, Mullard Space Science Laboratory, Holmbury St. Mary, RH5 6NT, UK.
             }

   \date{Received September 15, 1996; accepted March 16, 1997}
   \titlerunning{Magnetic Kelvin-Helmholtz Instability in the Solar Wind}
   \authorrunning{R. Kieokaew et al.}

 
  \abstract
   {The Kelvin-HeImholtz instability (KHI) is a nonlinear shear-driven instability that develops at the interface between shear flows in plasmas. KHI has been inferred in various astrophysical plasmas, and has been observed in situ at the magnetospheric boundaries of solar-system planets and through remote sensing at the boundaries of coronal mass ejections.}
   {KHI is also expected to develop at flow shear interfaces in the solar wind. While it was hypothesized to play an important role in the mixing of plasmas and in triggering solar wind fluctuations, its direct and unambiguous observation in the solar wind was still lacking. 
   }
   {We report in-situ observations of ongoing KHI in the solar wind using Solar Orbiter during its cruise phase. The KHI is found in a shear layer in the slow solar wind in the close vicinity of the Heliospheric Current Sheet, with properties satisfying linear theory for its development. Analysis is performed to derive the local configuration of the KHI. A 2-D MHD simulation is also set up with the empirical values to test the stability of the shear layer. In addition, magnetic spectra of the KHI event are analyzed. 
    }
   {We find that the observed conditions satisfy the KHI onset criterion from the linear theory analysis, and its development is further confirmed by the simulation. The current sheet geometry analyses are found to be consistent with KHI development. Additionally, we report observations of an ion jet consistent with magnetic reconnection at a compressed current sheet within the KHI interval. The KHI is found to excite magnetic and velocity fluctuations with power law scalings that approximately follow $k^{-5/3}$ and $k^{-2.8}$ in the inertial and dissipation ranges, respectively. Finally, we discuss reasons for the lack of in-situ KHI detection in past data.  
   }
   {These observations provide robust evidence of KHI development in the solar wind. This sheds new light on the process of shear-driven turbulence as mediated by the KHI with implications for the driving of solar wind fluctuations. 
   }

   \keywords{Solar-wind shear flows --
                Kelvin-Helmholtz instability --
                Slow solar-wind --
                Heliosphere --
                Space plasmas
               }

   \maketitle
%

\section{Introduction}
The magnetic Kelvin-Helmholtz instability (KHI) is a magnetohydrodynamic (MHD) shear-driven instability frequently observed in solar system plasmas. KHI can be induced at the surface between two media with different flow velocity and plasma conditions. This shear instability is fundamental and can be found in many flow shear systems throughout the Universe. The KHI that has been most studied in situ is at the Earth’s magnetopause and low-latitude boundary layers where periodic fluctuations of magnetic fields and plasma parameters are observed \citep[e.g.,][]{1981JGR....86..814H, 2000JGR...10521159F, Hasegawa2004, Foullon2008, Eriksson2016a}. This is because the Earth’s flank magnetopause is prone to the KHI onset condition which prefers weak magnetic field in the direction of the shear flow. Nevertheless, KHI has been observed in strong magnetic field environments such as in the solar corona \citep{Ofman2011} and at the flank of a coronal mass ejection (CME) \citep[e.g.,][]{Foullon2011a, Foullon2013, Mostl2013} via Extreme Ultraviolet (EUV) imaging by the Solar Dynamics Observatory (SDO). Variations of the KHI such as a sinusoidal mode have also been observed in a solar prominence \citep{Hillier2018} using Interface Region Imaging Spectrograph \citep{2014SoPh..289.2733D}.

Recently, remote observations above the solar corona using Heliospheric Imager instrument on board the Solar-Terrestrial Relations Observatory (STEREO/HI1) have revealed a transition in texture of the solar wind from highly anisotropic coronal plasma, or ``striae'', to more isotropic, or ``flocculated'', solar-wind plasma \citep{DeForest2016}. The transition is found to be consistent with the onset of hydrodynamic and MHD instabilities leading to development of turbulence. Qualitatively, this transition is found to occur near the first surface of plasma $\beta = 1$, where the $\beta$ changes from $\beta \ll 1$ near the Sun to $\beta \approx 1$, and the Alfvén critical surface where the solar wind speed reaches the Alfvén speed ($V = V_A$) \citep{Chhiber2018}. \citet{Ruffolo2020} propose that the transition from striae to flocculation of the young solar wind is powered by shear-driven instabilities such as the KHI, caused by the relative velocities of adjacent coronal magnetic flux tubes. This is supported by compressible MHD numerical simulations in which several features observed by Parker Solar Probe \citep[PSP;][]{2016SSRv..204....7F} including the magnetic ``switchback'' signatures near perihelia are reproduced. It is argued that the KHI can be triggered when the relative velocity between flux tubes is larger than the local Alfvén velocity ($\Delta V > V_A$) when the magnetic field is along the velocity shear, leading to shear-driven turbulence just outside the Alfvén critical zone. 

In theory, the KHI may develop at tangential discontinuities (TDs) because there are relative changes in velocity field ($\mathbf{V}$) and magnetic field ($\mathbf{B}$) across them and no normal magnetic field threading through them, which would otherwise stabilize the instability (as for rotational discontinuities). The solar wind is full of TDs \citep[e.g.,][]{1977JGR....82.3191B, Knetter2004, Neugebauer2010}, which separate different plasma regions. TDs are thought of as surfaces that separate adjacent solar wind flux-tubes \citep[e.g.,][]{Hollweg1982} that originated from granules, or meso/super granules  in the Sun's photosphere \citep[e.g.,][]{1986SoPh..107...11R, 1992sws..coll....1A} with different plasma properties and composition and spread out in the heliosphere \citep[e.g.,][]{2005LRSP....2....4B, 2008JGRA..113.8110B}. It was demonstrated theoretically by many authors that TDs can support MHD surface waves \citep[e.g.,][]{Hollweg1982}. In particular, TDs could support the KHI \citep[e.g.,][]{Burlaga1972, Neugebauer1986}. As for the flux tube picture, it was suggested that the KHI should be induced when adjacent flux tubes move relative to each other with a speed greater than the Alfvén speed \citep{Burlaga1972}. \citet{Zaqarashvili2014} consider the topology of magnetic flux tubes and found that, while the axial $\mathbf{B}$ of the flux tubes stabilizes the KHI, a slight twist in the $\mathbf{B}$ (i.e., as for when the $\mathbf{B}$ does not align with the $\mathbf{V}$) may allow the surface between them with $\Delta V < V_A$ to become unstable to the KHI.

Across TDs, an alignment between the change in velocity ($\Delta V$) and the change in magnetic field ($\Delta B$) are commonly observed \citep{Neugebauer1985, Neugebauer1986, Neugebauer2006, Knetter2004}. This alignment is not expected from the MHD discontinuity theory, unlike at rotational discontinuities where this alignment is expected \citep{Hudson1970}. This alignment of $\Delta V$ and $\Delta B$ is observed independent of the type of solar wind stream between 1 and 2.2 AU based on IMP 8 and Voyager 2 data \citep{Neugebauer1985}. Based on Helios data, \citet{Neugebauer1986} report that this alignment is also observed as close as $0.3$ AU. They further consider the possibility that the KHI may have developed across TDs and destroyed the random alignment between $\Delta V$ and $\Delta B$. Since the number of TDs per unit time decrease with distance from the Sun \citep[e.g.,][]{1979JGR....84.2773T, 1986JGR....91.8725L}, \citet{Neugebauer1986} suggested that the KHI may have destroyed TDs as the KHI growth rate becomes larger with decreasing Alfvén speed.

Despite all these postulations, the KHI has not been observed in situ in the solar wind to our knowledge. Here we report unambiguous in-situ KHI detection in Solar Orbiter observations \citep[SolO;][]{Muller2020} during its cruise phase. SolO is an ESA mission, launched on February 10, 2020, aimed to study the Sun and inner heliosphere from out-of-ecliptic vantage points. The cruise phase started in June 2020 with the in-situ instruments operated nominally after a successful commissioning. SolO was at $0.69$ AU during the observations presented here when several periodic fluctuations in plasma parameters are observed.

In this work, we report quasi-periodic magnetic and plasma variations within a shear layer in the slow solar wind consistent with the development of the KHI, supported by the linear theory analysis and numerical simulations. The paper is organized as follows. First, we present the instrumentation and overview, context, and KHI observations in Section~\ref{sec:obs}. Additionally, we report observations of magnetic reconnection signatures in the KHI. Then, we focus on the linear theory, boundary layer analyses, numerical simulations, and magnetic spectra of the KHI in Section~\ref{sec:results}. As this event shows clear evidence of the KHI in the solar wind, we discuss why it was not observed in past data, as well as its implications for solar wind dynamics in Section~\ref{sec:discussion}. We summarize our findings and discussion in Section~\ref{sec:conclusions}.

\section{Observations} \label{sec:obs}

\subsection{Instrumentation and overview}

   \begin{figure}
   \centering
   \includegraphics[width=3.25in]{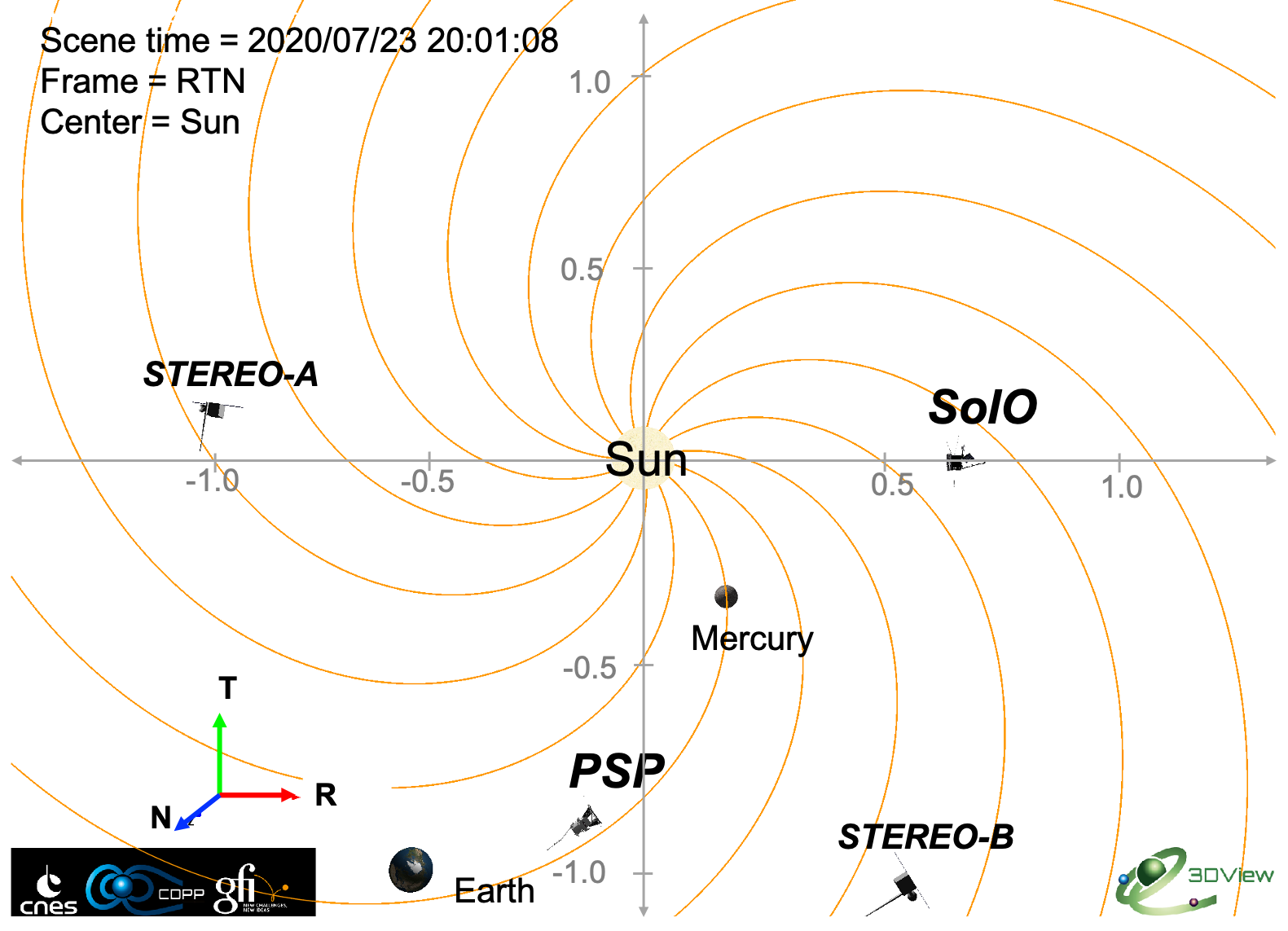}
      \caption{Overview of SolO and other spacecraft positions on 23 July 2020 at 20:01:08 UT in the RTN coordinates. Orange solid lines represent solar wind Parker spirals \citep{1958ApJ...128..664P} calculated using a constant velocity of $300$ km s$^{-1}$. This figure was obtained using the 3DView software \citep{GENOT2018111}.}
         \label{fig:SOpos}
   \end{figure}

We use magnetic field data from the fluxgate vector magnetometer \citep[MAG;][]{Horbury2020}. MAG continuously samples the magnetic field with the rate up to 16 vector/s in the normal mode and up to 128 vector/s in the burst mode with a precision of about 5 pT. We also use particle data from the Proton and Alpha Particle Sensor (PAS) that is part of the Solar Wind Analysis instrument suite \citep[SWA;][]{Owen2020}. PAS provides high-cadence measurements of 3D velocity distribution function of solar wind particles (electrons, protons, alpha particles, and heavy ions). We use the Radial Tangential Normal (RTN) coordinate system throughout this paper unless stated otherwise. In this system, the coordinates are centred at the spacecraft where $R$ is directed radially outward from the Sun to spacecraft, $T$ is longitudinal along the cross product of the Sun's rotation vector with $R$, and $N$ completes the right-handed orthogonal set, which points in the latitudinal direction.

   \begin{figure}[ht]
   \centering
   \includegraphics[width=3.5in]{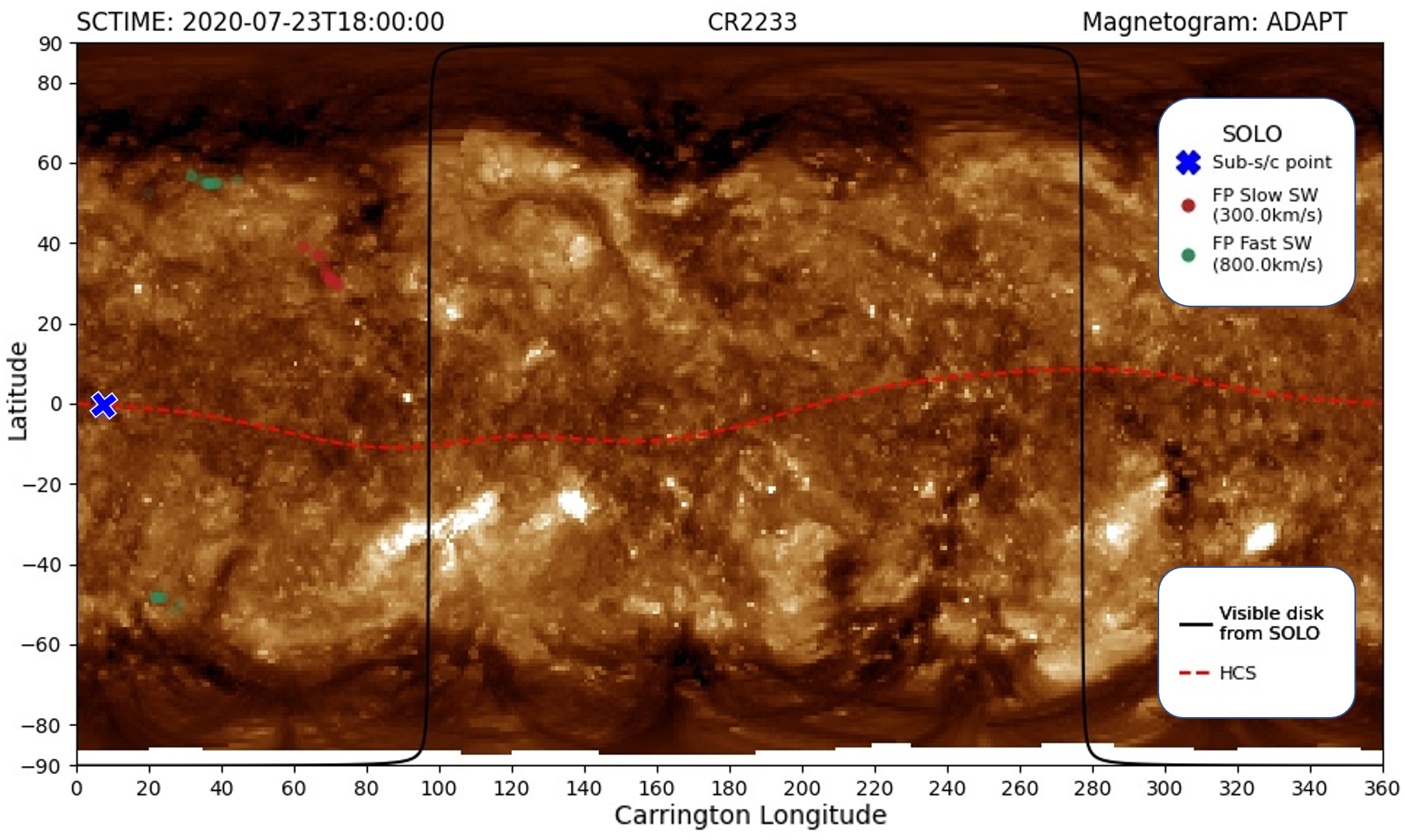}
      \caption{Connectivity map of the observations on July 23, 2020, produced from the Connectivity Tool \citep{Rouillard2020} to provide a global context. The background is a combined image from remote observation using SDO/AIA and STEREO-A/EUVI 193 $\AA$ Carrington maps. The SolO position is projected to the area of observation  and is shown as a blue cross (see text). The Heliospheric Current Sheet (HCS) is marked with a red dashed line. 
              }
         \label{fig:map}
   \end{figure}

On July 23 - 24, 2020, SolO was at the distance 0.69 AU from the Sun. From July 23 at 12:00 UT to July 24 at 12:00 UT, SolO was moving from the distance $r = 1.03 \times 10^8$ km ($0.68$ AU) to $r = 1.04 \times 10^8$ km ($0.70$ AU) from the Sun. Fig.~\ref{fig:SOpos} shows the projected position of SolO onto the equatorial plane with the solar wind Parker spiral (orange lines) \citep{1958ApJ...128..664P} calculated using a constant solar wind speed of $300$ km s$^{-1}$. This figure is obtained using the 3DView software \citep{GENOT2018111} available at \url{http://3dview.cdpp.eu/}. The radial speed of SolO was $V_R = 11.3$ km s$^{-1}$ throughout this interval. 

Fig.~\ref{fig:map} shows the global context on July 23, 2020 from 18 to 24 UT of the solar observations with the SolO position (blue cross) projected back to the Sun's surface. This figure is obtained from the Magnetic Connectivity Tool \citep{Rouillard2020}, accessible at \url{http://connect-tool.irap.omp.eu/}, that can be used to connect between remote observations and in-situ observations. The background image is produced by combining images taken by SDO and STEREO-A in EUV at $193$ $\AA$, using the Heliospheric Imagers (HI) instruments on board. The green and red dots indicate the points at the surface that are most likely to be magnetically connected to SolO, assuming uniform slow and fast wind flows. The SolO position is estimated to be at the blue cross position. The connected positions are projected on magnetograms from the ADAPT (Air Force Data Assimilative Photospheric Flux Transport) flux transport model \citep{2010AIPC.1216..343A, 2015SoPh..290.1105H}. On this image, the SolO position appears at low-latitude ($\sim 0^o$) and near the Heliospheric Current Sheet marked by a red dashed line. In this work, we report SolO observations of the KHI near an edge of the HCS.

   \begin{figure*}
   \centering
   \includegraphics[width=6.5in]{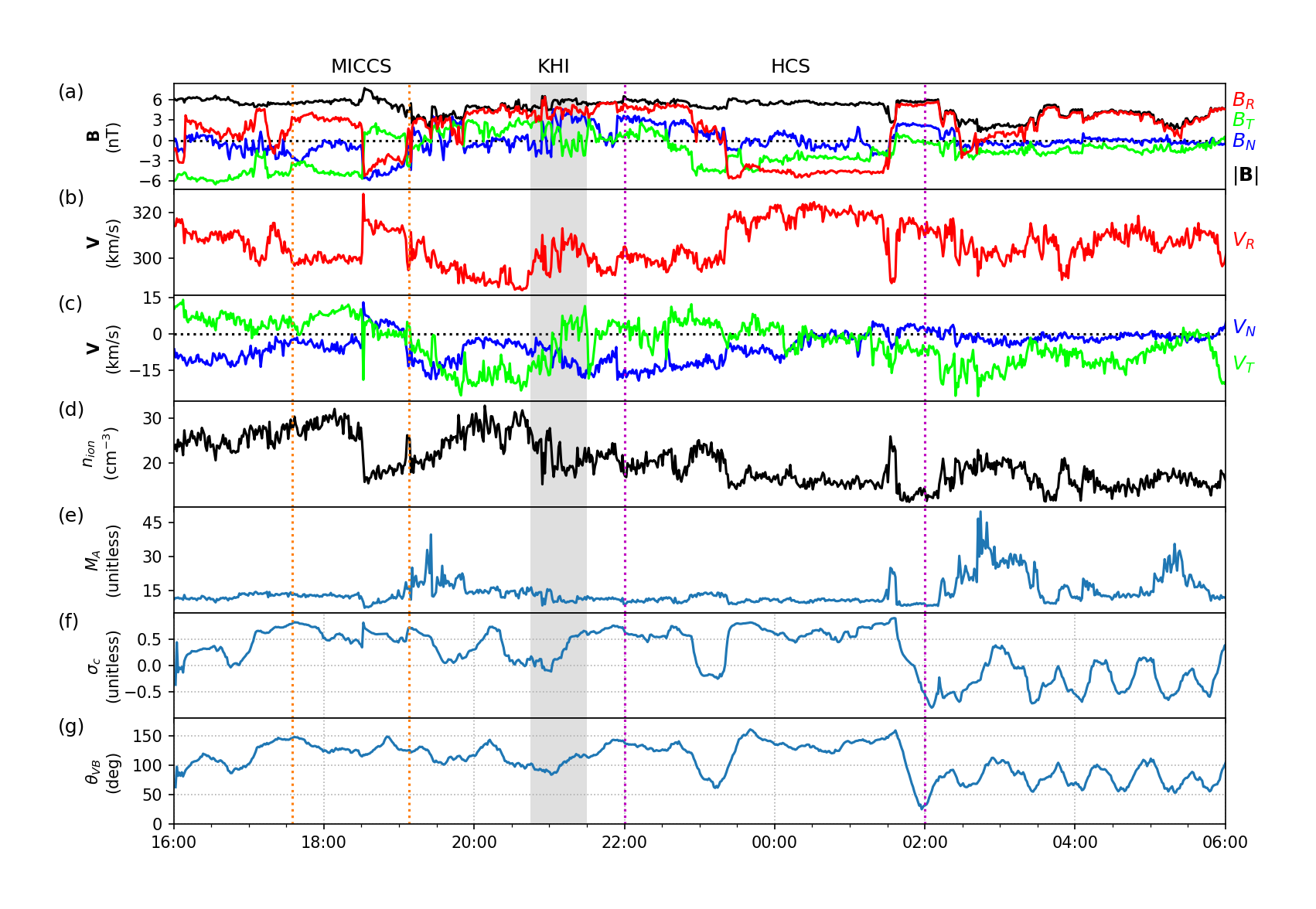}
   \caption{Solar Orbiter observations between 16:00 UT on 23 July 2020 and 6:00 UT on 24 July 2020 showing the context of the Kelvin-Helmholtz instability observation (shaded area). (a) Magnetic fields in the RTN coordinates. (b) Ion bulk velocity $V_R$ component. (c) Ion bulk velocity $V_T$ and $V_N$ components. (d) Ion number density. (e) Alfvén Mach number. (f) Normalized cross-helicity. (g) The angle between $\delta \mathbf{v}$ and $\delta \mathbf{b}$ (see text). The Heliospheric Current Sheet (HCS) is observed between 22:00 UT on 23 July 2020 and 2:00 UT on 24 July 2020, marked by purple dotted lines. The Magnetic Increase with Central Current Sheet (MICCS) is observed between 17:35 and 18:08 UT on 23 July 2020, marked by orange dotted lines.}
              \label{fig:12hr}
    \end{figure*}

\subsection{12-hour context}

Fig.~\ref{fig:12hr} shows SolO observations between 16:00 UT on July 23, 2020 and 6:00 UT on July 24, 2020, covering a 12-hour interval. In Fig.~\ref{fig:12hr}a, we show the magnetic field components and magnitude. Fig.~\ref{fig:12hr}b displays the radial component ($V_R$) and Fig.~\ref{fig:12hr}c the tangential ($V_T$) and out-of-ecliptic ($V_N$) components of the ion bulk flow velocity. Two main HCS crossings are observed between 22:00 UT on July 23 and 2:00 UT on July 24 in between the purple vertical dotted lines. The HCS is characterized by a large-scale ($\sim$several hours) change in the polarity of the radial magnetic field ($B_R$). Around 18:00 UT on July 23, we mark a meso-scale structure ($\sim$a few hours) between vertical orange dotted lines, characterizing bipolar magnetic fields with a sharp magnetic transition (i.e., current sheet) at the center. At this central current sheet, there is an ion jet clearly seen in $V_R$ and $V_N$, probably indicating ongoing magnetic reconnection. This structure is known as Magnetic Increase with Central Current Sheet (MICCS) and it has been observed with Parker Solar Probe \citep{Fargette2021}; this signature corresponds to two magnetic flux tubes that become interlinked and with magnetic reconnection at the interface \citep[e.g.,][]{Louarn2004, Kacem2018, Oieroset2019, Kieokaew2020}. Although mentioned here for context, this feature is not discussed any more in our study.

       \begin{figure*}
   \centering
    \includegraphics[width=6.5in]{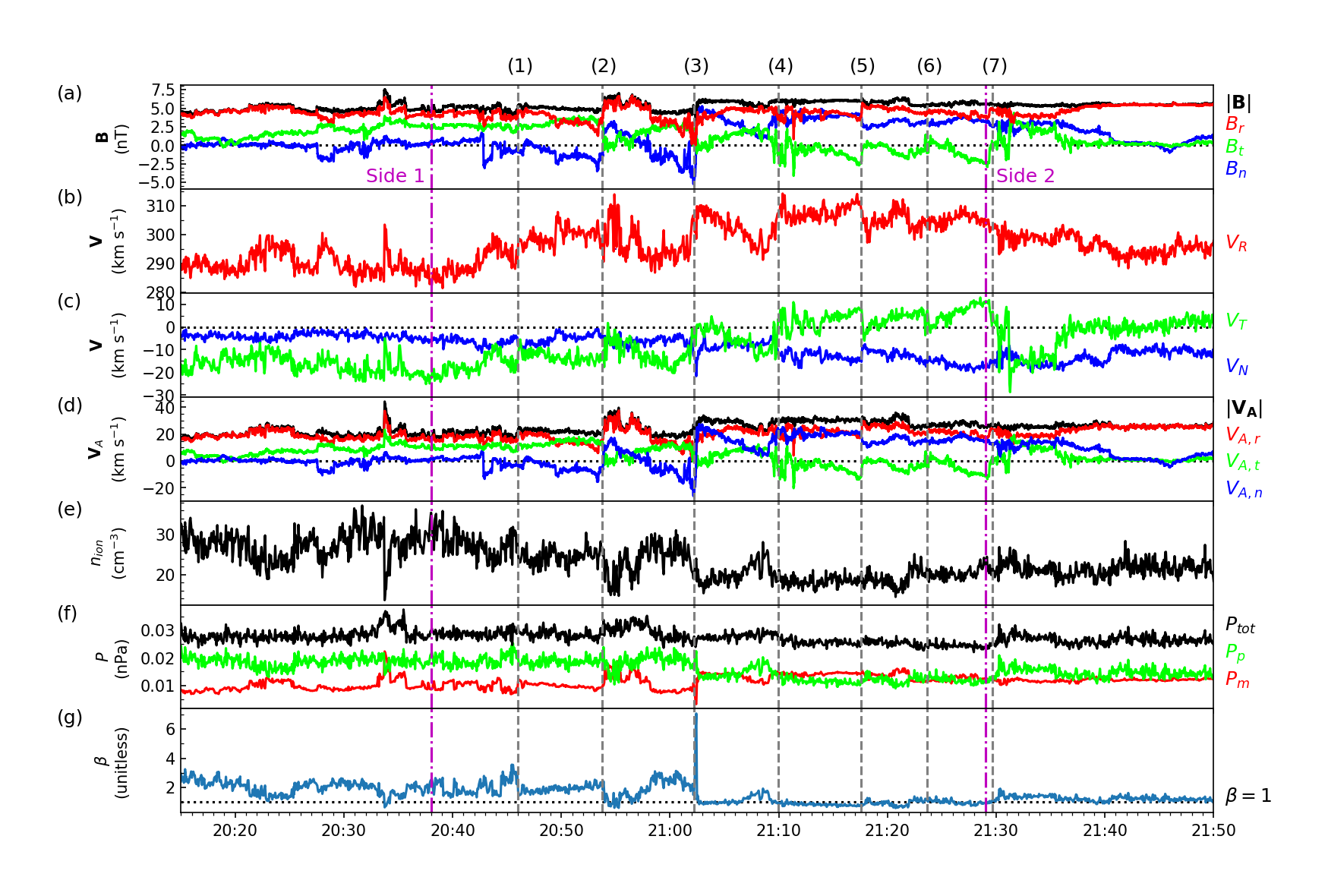}
   \caption{Solar Orbiter observations on 23 July 2020 between 20:10 UT and 21:50 UT. (a) Magnetic field in the RTN coordinates. (b) Ion bulk velocity $V_R$ component. (c) Ion bulk velocity $V_T$ and $V_N$ components. (d) Alfvén speed $|\mathbf{V}_A|$ and Alfvén velocity components ($V_{A,r}, V_{A,t}, V_{A,n}$). (e) Ion number density. (f) Plasma pressure $P_p$, magnetic pressure $P_m$, and total pressure $P_{tot}$. (g) Plasma beta $\beta$. The KH waves can be noticed between 20:45 and 21:30 UT. The vertical grey dashed lines (1) - (7) mark the boundary layer crossings corresponding to KH trailing edges. The magenta dashed-dotted lines mark times when the plasma parameters reach asymptotic values on Side 1 (20:38 UT) and Side 2 (21:29 UT) of the shear layer. }
              \label{fig:KH}
    \end{figure*}

From 20:45 to 21:30 UT on July 23 in Fig.~\ref{fig:12hr}c, we observe a velocity shear of about $30$ km s$^{-1}$ in the $V_T$ (green) and about $20$ km~s$^{-1}$ in the $V_N$ (blue) components. This velocity shear interval features periodic fluctuations in the magnetic fields (Fig.~\ref{fig:12hr}a) and the radial velocity ($V_R$; Fig.~\ref{fig:12hr}b). The shear layer is also accompanied by changes in the ion number density shown in Fig.~\ref{fig:12hr}d. This interval as shaded in grey exhibits KH-like signatures; we show a zoom-in of this interval in the next section. 

Figs.~\ref{fig:12hr}e, ~\ref{fig:12hr}f, and ~\ref{fig:12hr}g show the Alfvén Mach number ($M_A$), the normalized cross-helicity ($\sigma_c$), and the angle between $\mathbf{V}$ and $\mathbf{B}$ changes ($\theta_{VB}$), respectively. During the shaded interval, the $M_A$ is $12$, indicating that the solar wind is super-Alfvénic. The $\sigma_c$ \citep{1982JGR....87.6011M, Roberts1992} is calculated from $\sigma_c = 2 \langle \delta \mathbf{v} \cdot \delta \mathbf{b} \rangle / \langle | \delta \mathbf{v}|^2 + |\delta \mathbf{b}|^2 \rangle$, where $\delta \mathbf{v} = \mathbf{V} - \langle \mathbf{V} \rangle_{20min}$ and $\delta \mathbf{b} - \langle \mathbf{b} \rangle_{20min}$ are the velocity field and magnetic field fluctuations from the $20-$min running averages of $\mathbf{V}$ and $\mathbf{b} = \mathbf{B}/\sqrt{\mu_0 \rho}$, where $\mu_0$ is the vacuum permeability and $\rho$ is the average proton mass density, respectively. The magnetic field $\mathbf{b}$ is measured in Alfvén speed units in km s$^{-1}$. The sum ($\langle \rangle$) brackets are also taken over $20$-min running averages. $\sigma_c$ relates to the cross helicity $H_c =  \frac{1}{2} \int (\mathbf{V} \cdot \mathbf{b})  d^2 x $ and the energy per unit mass $E = \frac{1}{2} \int (\mathbf{V}^2 + \mathbf{b}^2) d^2 x$ through the relation $\sigma_c = 2 H_c / E$. This quantity measures the Alfvénicity with $\sigma_c = \pm 1$ indicating Alfvénic fluctuations, where the signs $+, -$ correspond to Alfvénic propagation anti-parallel or parallel to the mean field, respectively. During the shaded interval, $\sigma_c$ is fluctuating around $0.5$, indicating that the solar wind is not strongly Alfvénic. The angle $\theta_{VB}$ is calculated from $ \arccos (\delta \mathbf{v} \cdot \delta \mathbf{b} / (|\delta \mathbf{v}||\delta \mathbf{b}|))$. For the shaded time period, Fig.~\ref{fig:12hr}g shows that the angle between $\delta \mathbf{v}$ and $\delta \mathbf{b}$ is spreading from $90^o$ to $120^o$. The non-alignment of $\delta \mathbf{v}$ and $\delta \mathbf{b}$ during this interval indicates the possibility for this shear layer to be unstable to the KHI. The KHI would be suppressed if the magnetic field change aligns with the velocity field change due to the magnetic tension exerted in the direction of the shear flow \citep[e.g.,][]{1961hhs..book.....C}, i.e., when $\theta_{VB}$ is near $0^o$ or $180^o$.

 \subsection{KHI observations} \label{subsec:KHI}
 
Fig.~\ref{fig:KH} shows a zoom-in around the shaded time period in Fig.~\ref{fig:12hr} between 20:45 and 21:30 UT on July 23, 2020. The quasi-periodic fluctuations in several parameters resemble in-situ KH waves at the interface between shear flows \citep[e.g.,][]{Hasegawa2004}. The magnetic field in Fig.~\ref{fig:KH}a clearly shows repeated, homologous fluctuations in all components. We mark sharp magnetic rotations (1) - (7) with vertical dashed lines. Fig.~\ref{fig:KH}b shows $V_R$ and Fig.~\ref{fig:KH}c shows $V_T$ and $V_N$ components of the ion bulk velocity. The velocity shear is clearly seen in the $V_T$ component. To mark the shear layer, we define Side 1 at 20:38 UT as the beginning of the shear interval and Side 2 as the end of the shear interval at 21:29 UT, marked by magenta dotted-dashed lines. Sides 1 and 2 are characterized mainly by the tangential velocity $V_T$ which reached local minimum and maximum, respectively. The periodic features are also seen in the velocity fields especially for the tangential ($V_T$) component in Fig.~\ref{fig:KH}c in which sudden, sharp transitions colocate with the changes in the magnetic field marked by the vertical dashed lines. To facilitate discussion, we define the magnetic rotations (1) - (7) as ``wave edges'' that mark sudden changes in both magnetic and velocity fields. Table~\ref{table:waveedge} notes times of these wave edges and time differences between these edges. These time differences correspond to the period of the waves. The average time difference between wave edges is $7$ minutes $17$ seconds (s) and the standard deviation is $55$ s. We analyze the wave edges in detail in Section~\ref{subsec:boundary-analysis}. 

\begin{table}
\caption{Timing of the KH wave edges marked in Fig.~\ref{fig:KH} and time difference between them.}              
\label{table:waveedge}      
\centering                                      
\begin{tabular}{ccc}          
\hline\hline                        
Numbers & Times (UT) & Time difference (mm:ss) \\    
\hline                                   
    (1)         &         20:45:59 &  - \\  
    (2)       &        20:53:43  & 07:44 \\  
   (3)        &       21:02:11  &  08:28 \\            
   (4)        &        21:09:56  & 07:45 \\  
    (5)        &        21:17:32  & 07:36 \\  
    (6)       &        21:23:39 & 06:07 \\      
   (7)       &        21:29:38 & 05:59 \\
 \hline\hline
\end{tabular}
\end{table}

Fig.~\ref{fig:KH}d shows Alfvén speed and Alfvén velocity components ($\mathbf{V}_A = \mathbf{B} /\sqrt{\mu_0 n_i m_p}$, where $n_i$ is the ion number density and $m_p$ is the proton mass). The average Alfvén speed in this interval is $26$ km s$^{-1}$. Fig.~\ref{fig:KH}e shows the ion number density ($n_{ion}$). Despite some fluctuations, $n_{ion}$ gradually changes from $30$ cm$^{-1}$ at Side 1 to $22$ cm$^{-1}$ at Side 2. Fig~\ref{fig:KH}f shows the magnetic ($P_m$), thermal ($P_p$), and total pressures ($P_{tot} = P_m + P_p$). The total pressure is approximately constant, indicating approximate pressure balance across the shear layer. Fig.~\ref{fig:KH}g shows the ion beta ($\beta$). The $\beta$ is changing from $\beta \approx 2$ at Side 1 to $\beta \approx 1$ at Side 2. A strong peak in $\beta$ of about $\beta \approx 7$ is observed adjacent to wave edge (3). At wave edge (3), there is an ion jet in $V_N$ (blue in Fig.~\ref{fig:KH}c) with $\Delta V_N \sim 11$ km s$^{-1}$ that colocates with the magnetic rotation observed in $B_N$ (blue in Fig.~\ref{fig:KH}a). We explain that this ion jet may be produced by magnetic reconnection, in Section~\ref{subsec:recon}.

Table~\ref{table:values} summarizes the values of $\mathbf{V}, \mathbf{B}$ and $n_{ion}$ at Sides 1 and 2. Considering the velocity change across the shear layer, we define $\Delta \mathbf{V} = \mathbf{V}_2 - \mathbf{V}_1$, where the subscripts $1$ and $2$ label Sides 1 and 2, respectively. The velocity change is $\Delta \mathbf{V} = (\Delta V_R, \Delta V_T, \Delta V_N) = (30, 30, -11) $ km s$^{-1}$, with $|\Delta \mathbf{V}| = 44$ km s$^{-1}$. The ratio of the velocity change across the shear layer to the average Alfvén speed is therefore $\Delta V / V_A = 1.7$. An important criterion of the KHI from the linear theory of stability of a shear layer in plasmas considered by \citet{1961hhs..book.....C},  \citet{Sen1963}, and \citet{Landau1960} is that the KHI would be suppressed for $\Delta V < {V_A}$, i.e., when the velocity shear is less than the local Alfvénic speed. Note that this condition is obtained in a simplistic configuration of an infinitely thin shear layer embedded in the uniform density and magnetic field and with the magnetic field being exactly along the velocity shear. The observed shear velocity exceeds the local Alfvén speed for this approximate consideration and thus the KHI is not suppressed. In Section~\ref{subsec:linear-theory}, to test whether the observed local shear conditions support the KHI growth, we consider the KHI onset condition by taking into account the nonuniform conditions on either side of the shear layer derived from the linear theory. We also set up a local MHD simulation using these empirical conditions in Section~\ref{subsec:sim}.

\begin{table}
\caption{Parameter values on Side 1 and Side 2 of the shear layer marked in Fig.~\ref{fig:KH}.}              
\label{table:values}      
\centering                                      
\begin{tabular}{c r r c}          
\hline\hline                        
Parameters & Side 1 & Side 2  & Units \\    
\hline                                   
    $V_R$         &        $285$         &        $305$      &    km s$^{-1}$ \\  
    $V_T$        &        $-20$         &        $10$         &        km s$^{-1}$   \\  
    $V_N$        &       $-5$           &         $-16$       &    km s$^{-1}$    \\      
\hline       
    $B_R$        &        $4$          &           $4$         &       nT   \\  
    $B_T$        &        $2$         &          $-2$          &      nT  \\  
    $B_N$       &        $0$          &           $3$         &       nT  \\      
\hline                                          
   $n_{ion}$        &        $30$          &           $22$         &       cm$^{-3}$ \\
 \hline\hline
\end{tabular}
\end{table}

\subsection{Magnetic reconnection} \label{subsec:recon}

   \begin{figure}[ht]
   \centering
   \includegraphics[width=3.5in]{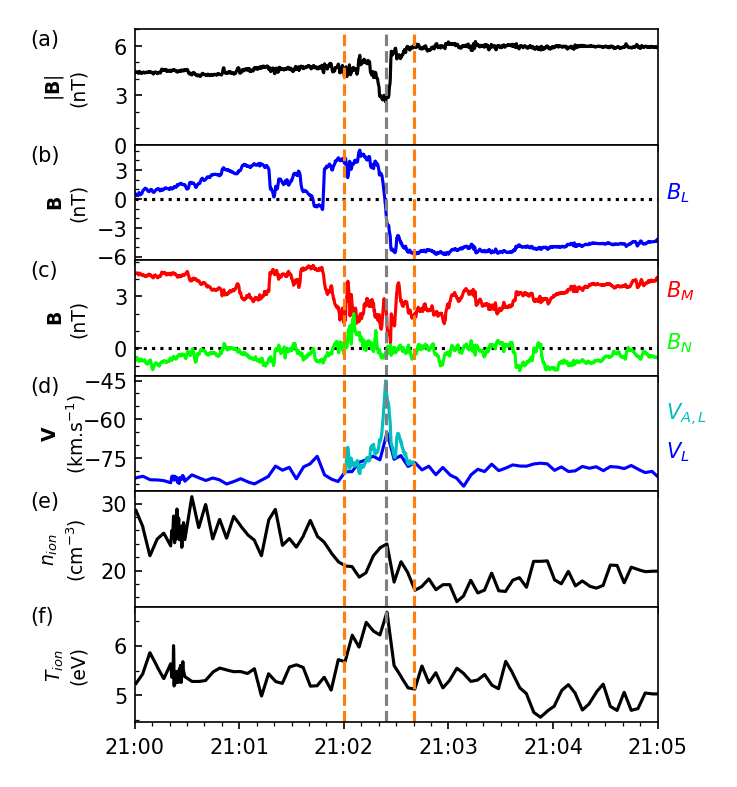}
      \caption{Magnetic reconnection signatures at wave edge (3) in Fig.~\ref{fig:KH}. (a) Magnetic field strength. (b) Magnetic field reconnecting $L$-component. (c) Magnetic field $M, N$ components. (d) Ion bulk velocity $V_L$ component with the predicted velocity jet ($V_{A,L}$; cyan). (e)  Ion number density. (f) Ion temperature. Dashed orange lines mark the extent of the ion jet. A dashed grey line marks the maximum ion velocity during the jet.
              }
         \label{fig:recon}
   \end{figure}

Fig.~\ref{fig:recon} shows a zoom-in at wave edge (3) in Fig.~\ref{fig:KH} that contains an ion jet colocated with the change in magnetic field seen in the $N$ component (blue). The current sheet interval at the wave edge is delineated by orange dashed lines between 21:02:00 and 21:02:40 UT. The magnetic field magnitude in Fig.~\ref{fig:recon}a shows a drop at the centre of the current sheet of about $2.5$ nT. To clearly see the ion jet, we transform the magnetic field into local current sheet $LMN$ coordinates using the hybrid Maximum Variance Analysis (MVA) technique \citep{Gosling2013}. In this coordinate system, $L$ points in the magnetic shear direction (i.e., the reconnecting component), $N$ points in the direction normal to the shear plane, and $M = N \times L$ points in the out-of-plane direction (i.e., the guide-field direction). We transform the magnetic field from the $RTN$ coordinates to $LMN$ coordinates as follows. First, the current sheet normal $N$ is obtained from the cross-product of the asymptotic $10$-s averaged magnetic fields just outside the current sheet interval. Second, the $M$ component is obtained from $N \times L_1$, where $L_1$ is the maximum variance direction obtained from the MVA technique \citep{Sonnerup1967} applied in the current sheet interval. Finally, $L= M \times N$ completes the right-handed orthonormal system. We obtain $L = [-0.285, 0.245, -0.924], M = [0.882, 0.44, -0.156] $, and $N = [0.37, -0.863, -0.343]$.

The magnetic field rotation is clearly seen in the $B_L$ component in Fig.~\ref{fig:recon}b. $B_L$ rotates from positive to negative during the current sheet interval in colocation with the ion jet seen in the $V_L$ component in Fig.~\ref{fig:recon}e. Note that the velocity data have lower cadence ($4$ s) than the magnetic field data ($1$ s), thus the velocity is sampled only at a few points in the vicinity of the current sheet interval. The velocity peak is marked by a grey dashed line. This ion jet has a magnitude of $\Delta V = 13$ km s$^{-1}$. Although there is only one velocity measurement associated with the peak inside the jet, one should note that the operation cycle of the PAS instrument is such that each sample is made over only $1$ s every $4$ s. The measurement is thus made over a limited time exactly in the center of the current sheet, rather than over $4$ s which would have led to significant time aliasing across the current sheet.

The change in $B_L$  correlates with the change in $V_L$ negatively on the inbound side and positively on the outbound side. This sequence of correlations is consistent with a jet that is produced by magnetic reconnection. Although we acknowledge that there is only one velocity measurement in the jet, there are additional signatures that are consistent with the interpretation of magnetic reconnection, such as the enhanced ion number density shown in Fig.~\ref{fig:recon}e, consistent with mixing of ions from either side of the current sheet in the reconnection exhaust \citep{Gosling2005} and the enhanced ion temperature consistent with plasma heating within the reconnection exhaust \citep{2014GeoRL..41.7002P} in Fig.~\ref{fig:recon}f.

To more quantitatively assess whether this jet is consistent with reconnection, we consider the Walén relation: $\Delta \mathbf{V}_A \sim \pm \Delta \mathbf{B} / (\mu_0 m_p n_{ion})^{1/2}$, where $+$ or $-$ is applied for a positive or negative correlation between $\mathbf{B}$ and $\mathbf{V}$, respectively, \citep{Hudson1970, Paschmann1986} within the exhaust as bounded by dashed orange lines in Fig.~\ref{fig:recon}. The predicted jet velocity is shown as cyan in Fig.~\ref{fig:recon}d. The predicted velocity produces a trend that resembles the observed jet. However, the predicted velocity jet from the Walén relation is estimated to be $\Delta V_A = 33$ km s$^{-1}$. Thus, the observed jet has a velocity that is $40$\% of the predicted jet velocity. A sub-Alfvénic reconnection jet is not unusual in observations. In the literature, sub-Alfvénic jet speeds are found when the spacecraft crosses the reconnection exhaust near the X-line. The development of secondary instabilities due to high plasma-$\beta$ was also found to lower the reconnection jet speed \citep{Haggerty2018}.  The presence of reconnection within KH waves may imply that it is produced as a consequence of vortex-induced-reconnection \citep[e.g.,][]{Nykyri2001, Nakamura2006, Karimabadi2013, Eriksson2016a}, which can be triggered when KH vortices develop and create thin current sheets between them. We further discuss this possibility in the discussion section.

\section{Results} \label{sec:results}

\subsection{Linear theory analysis} \label{subsec:linear-theory}

To test whether the observed local conditions in Table~\ref{table:values} satisfy the KHI onset criterion, we consider the stability of a shear layer derived using the linear theory of \citet{1961hhs..book.....C}. The KHI onset criterion derived for an infinitely thin boundary layer in an incompressible plasma can be written \citep{1975pine.book.....H} as 
   \begin{equation}
      \left[ \mathbf{k} \cdot (\mathbf{V}_1 - \mathbf{V}_2) \right]^2 > \frac{n_1 + n_2}{\mu_0 m_p n_1 n_2} \left[ (\mathbf{k} \cdot \mathbf{B}_1)^2 + (\mathbf{k} \cdot \mathbf{B}_2)^2 \right] \label{eq:onset}
    \end{equation}
where $\mathbf{k}$ is the wave vector, $\mathbf{V}$ is the velocity field, $\mathbf{B}$ is the magnetic field, and $n$ is the ion number density, with the subscripts $1, 2$ representing either side of the shear layer. The phase velocity of the KH mode, associated with the real part of the KH dispersion relation $V_{ph} = \omega / k$, where $\omega$ is the wave frequency and $k$ is the wave number, is given as 
   \begin{equation}
V_{ph} =  \dfrac{n_1 \mathbf{k} \cdot \mathbf{V}_1 + n_2 \mathbf{k} \cdot \mathbf{V}_2}{k (n_1 + n_2)} \label{eq:Vph}
   \end{equation}
The growth rate of the KHI, associated with the imaginary part of the dispersion relation, can be written as 
\begin{equation}
\gamma = [ \alpha_1 \alpha_2 [ (\mathbf{V}_1 - \mathbf{V}_2) \cdot \mathbf{k}]^2 - \alpha_1 ( \mathbf{V_{A,1}} \cdot \mathbf{k})^2 -  \alpha_2 ( \mathbf{V_{A,2}} \cdot \mathbf{k})^2 ]^{1/2} \label{eq:growth}
\end{equation}
where $\alpha_1 = n_1 / (n_1 + n_2)$ and $\alpha_2 = n_2 / (n_1 + n_2)$, and $\mathbf{V_{A,1}}, \mathbf{V_{A,2}}$ label the Alfvén speeds on either side of the boundary. 

   \begin{figure}
   \centering
   \includegraphics[width=3.5in]{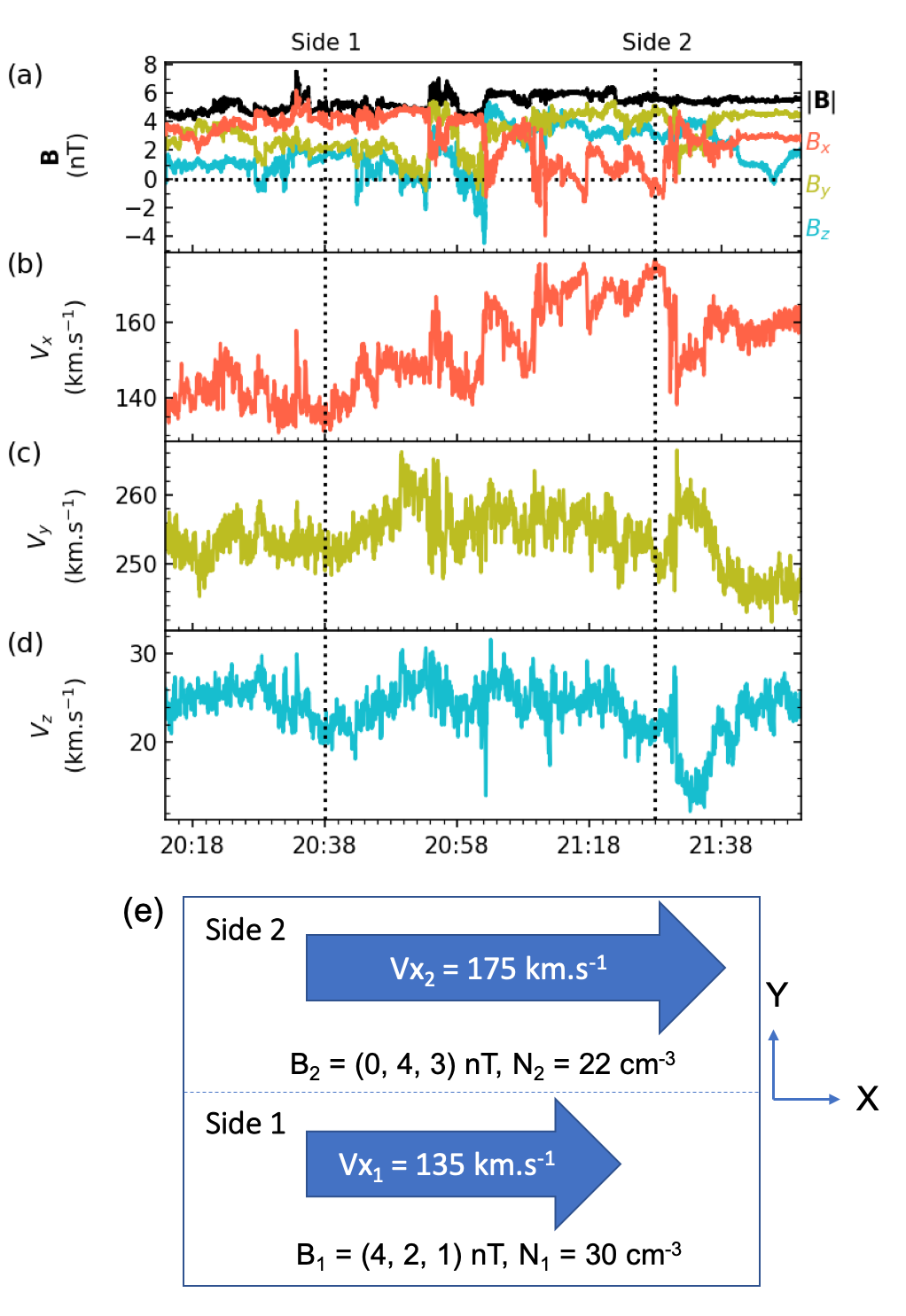}
      \caption{Magnetic and velocity fields in the local shear frame obtained from the application of the MVA to the ion bulk velocity. (a) Magnetic field in the maximum ($X$), intermediate ($Y$), and minimum ($Z$) variance directions and its magnitude. (b, c, d) Velocity field in the $X$, $Y$, and $Z$ directions, respectively. (e) Simplified shear boundary layer configuration obtained from the transformation to the maximum shear frame. 
              }
         \label{fig:MVA-box}
   \end{figure}

To simplify the configuration of the observed shear layer, we transform the velocity field using the application of the MVA technique to the ion bulk velocity from 20:14 to 21:50 UT. The maximum, intermediate, and minimum variance directions are found to be $[0.53, 0.79, -0.32], [0.84, -0.53, 0.07]$, and $[0.12, 0.31, 0.95]$, respectively. The ratios of the maximum to the intermediate eigenvalues ($\lambda_1 / \lambda_2$) and the intermediate to the minimum eigenvalues ($\lambda_3 / \lambda_2$) are $7.7$ and $2.0$, respectively, indicating reliable estimations \citep{1972JGR....77.1321S}. The maximum variance direction is the direction where we find the velocity jump (i.e., the wave amplitude direction); it is assigned as $Y$. The minimum variance direction is the invariant direction; it is assigned as $Z$. The intermediate variance direction is the wave propagation direction; it is assigned as $X$. The transformed $\mathbf{B}$ and $\mathbf{V}$ are shown in Figs.~\ref{fig:MVA-box}a - \ref{fig:MVA-box}d. The velocity jump is clearly seen in  Fig.~\ref{fig:MVA-box}b with $\Delta V = 40$ km s$^{-1}$. Fig.~\ref{fig:MVA-box}e shows a simplified configuration of this shear layer in the $X-Y$ plane. Note that the $V_y$ and $V_z$ are nearly constant and thus not shown in this figure.

We may calculate a KH growth rate from the simplified configuration in Fig.~\ref{fig:MVA-box}e. In this frame, the shear velocity is $\Delta \mathbf{V} = \mathbf{V}_2 - \mathbf{V}_1 \approx (40, 0, 0) $ km s$^{-1}$. Since the magnetic field perpendicular to the shear direction, i.e., the $B_z$ component (the invariant direction), does not impact the KHI \citep{1961hhs..book.....C}, we can ignore this component. Assuming that the wave vector $\mathbf{k}$ is in the $X-Y$ plane and makes an angle $\phi$ from the $Y$-direction such that $\mathbf{k} = (k \cos \phi, k \sin \phi, 0)$, Eq.~\ref{eq:growth} can be written as 
\begin{eqnarray}
\left( \frac{\gamma}{k} \right)^2 &=& \frac{\rho_1 \rho_2}{(\rho_1 + \rho_2)^2} \left[ \Delta V_x \cos \phi + \Delta V_y \sin \phi \right]^2 \nonumber \\
& &  - \frac{1}{\mu_0 (\rho_1 + \rho_2)} \left[ B_{1,x} \cos \phi + B_{1,y} \sin \phi \right]^2 \nonumber \\
& &  - \frac{1}{\mu_0 (\rho_1 + \rho_2)} \left[ B_{2,x} \cos \phi + B_{2,y} \sin \phi \right]^2  \label{eq:growthxy}
\end{eqnarray}
where $\rho_i = m_p n_i$, $i = 1, 2$. We find positive growth rates for an arbitrary angle $\phi$ with a maximum growth rate ($\gamma / k$) of $16$ km s$^{-1}$. This means that the observed conditions of this event are unstable to the KHI. In brief, the linear theory analysis supports the KHI interpretation.

 Assuming that the wave vector $\mathbf{k}$ is in the same direction as the flow, we obtain the KH phase velocity from Eq. (\ref{eq:Vph}) to be $V_{ph} = 152$ km s$^{-1}$. Since the average wave period from Table~\ref{table:waveedge} is found to be $437 \pm 55$ s ($7.3 \pm 0.9$ minutes), the KH wavelength is estimated to be $\lambda_{KH} = 66,400 \pm 8,400$ km or $0.10 \pm 0.01$ solar radii. The linear theory analysis of a finite-thickness shear layer by \cite{Miura1982} predicts that the fastest growing mode occurs for $k  L \sim 0.5 - 1.0$, where $ L$ is the initial shear layer thickness. Using $k = 2 \pi / \lambda$, the fastest growing mode should have the wavelength of $2 \pi L - 4 \pi L$. Using our estimated $\lambda_{KH}$, we estimate the initial shear layer thickness to be $L \approx 5,300$ - $10,500$ km. Note that this estimate ignores any vortex merging or potential influence of pre-existing turbulence that could influence the size of the vortices that are observed.

\subsection{KH simulation} \label{subsec:sim}

   \begin{figure*}
   \centering
    \includegraphics[width=6.5in]{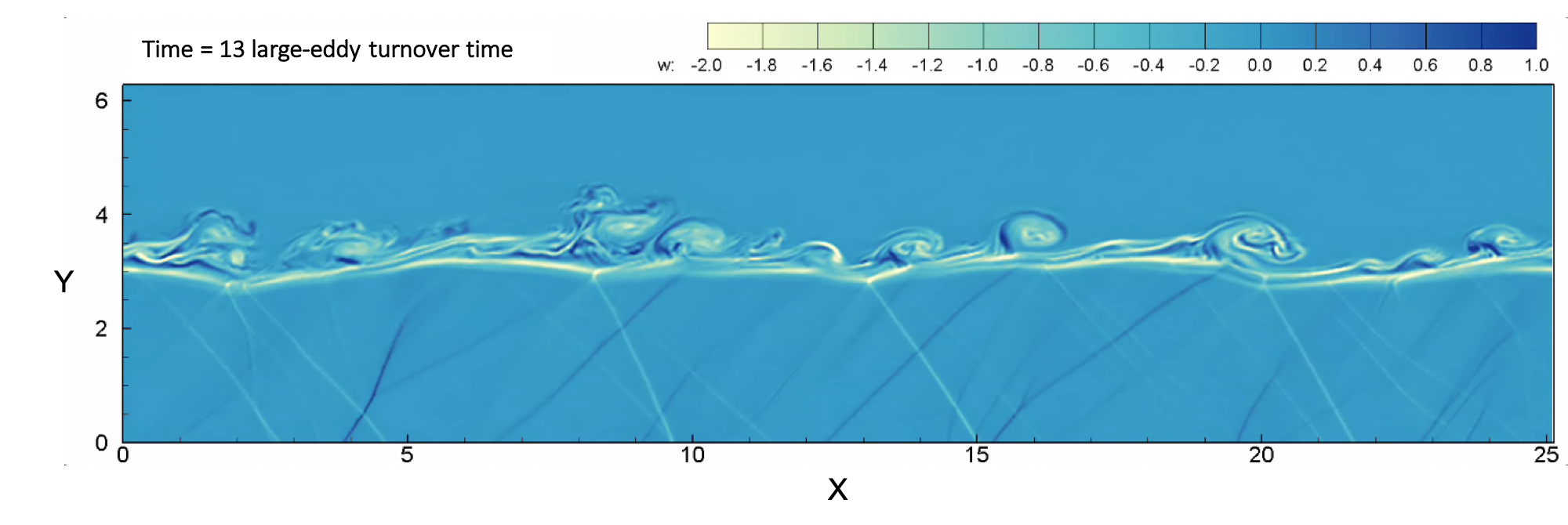}
   \caption{A snapshot of the numerical simulation of the KHI using empirical values of the SolO event from Side 1 and Side 2. The color-scale represents values of the out-of-plane flow vorticity ($\omega$). The KHI quickly reaches the non-linear stage where rolled-up KH vortices form and coalesce. The stripes in vorticity in the lower part of the simulation are shocks produced by the supersonic flow as the Mach number $\sim 3$ (see text).}
              \label{fig:KHsim}%
    \end{figure*}

To further test whether the observed conditions would support the KHI, we performed a numerical simulation using SolO observations for Side 1 and Side 2 as boundary conditions. We exploit a numerical simulation that solves compressible MHD equations via a hybrid compact-weighted essentially non-oscillatory (WENO) scheme \citep{2016JCoPh.306...73Y}. This hybrid scheme couples a sixth-order compact finite difference scheme for smooth regions and a fifth-order WENO scheme in shock regions, suitable for capturing strong discontinuities in MHD systems. The time stepping is performed by the third-order Runge-Kutta scheme. This code has been used to study compressible MHD turbulence \citep{2016PhRvE..93f1102Y, 2017PhFl...29c5105Y} and shear-driven turbulence by the KHI near the Sun \citep{Ruffolo2020}. 

To simulate our event, we consider the local shear frame in Fig.~\ref{fig:MVA-box}e, derived in Section~\ref{subsec:linear-theory}. Moreover, we consider the local KHI frame that travels with the KH phase speed at $V_{ph} \approx 150$ km s$^{-1}$. In this frame, the speeds on Sides 1 and 2 are $U_{X,1} = -15$  km s$^{-1}$ and $U_{X,2} = 25$ km s$^{-1}$, respectively. Since the magnetic field perpendicular to the shear flow does not impact the KH growth \citep{1961hhs..book.....C}, we only include magnetic field in the shear flow direction. The ion number density values on Sides 1 and 2 are $n_1 = 30$ cm$^{-3}$ and $n_2 = 22$ cm$^{-3}$, respectively. The ion temperature values on Sides 1 and 2 are set to $T_1 = 1.32$ eV and $T_2 = 3.63$ eV, respectively. The ion $\beta$ is set to $1$. The magnetosonic Mach number across the shear layer is $\Delta U / c_{s,1} = 2.75$, where $\Delta U = 40$ km s$^{-1}$ and $c_{s,1}$ is the sound speed on Side 1. The Alfvén Mach number is $\Delta U / V_{A,1} = 2.52$, where $V_{A,1}$ is the Alfvén speed on Side 1. 

The numerical simulation is performed using a $L_x \times L_y = 8\pi \times 4\pi$ domain with $n_x \times n_y = 1024 \times 512$ resolution with periodic boundary conditions in the $X$-direction. For simplicity, equal viscosity and resistivity $\mu = \eta$ are used, i.e., the magnetic Prandtl number is set to unity. We solve the dimensionless form of the MHD equations by introducing several reference scales. The normalizations are $U_0 = 100$ km s$^{-1}$, $n_0 = 30$ cm$^{-3}$, $B_0 = 25$ nT, and $T_0 = 66$ eV. The simulation is 2-D as we ignore the invariant direction and only impose the magnetic field in the direction of the shear flow. 

We set up double shear layers in the simulation domain similar to those of \citet{Ruffolo2020}. The velocity and magnetic profiles are only set in the $X$-direction and both are colocated. The velocity profile is given by
\begin{eqnarray}
u_x = U_\alpha \left[ 1 - \tanh \left( \frac{y - L_y / 4}{d} \right) + \tanh \left( \frac{y - 3L_y / 4}{d} \right) \right] + U_\beta, \label{eq:Uy}
\end{eqnarray}
where $U_\alpha = \left( \frac{U_1 - U_2}{2} \right)$, $U_\beta = \left( \frac{U_1 + U_2}{2} \right) $, $U_1 = -0.15 U_0$, $U_2 = 0.25 U_0$, and $d = 0.003 L_y$ is the half thickness of the shear layer. The magnetic profile is given in a similar way as 
\begin{eqnarray}
B_x = B_\alpha \left[ 1 - \tanh \left( \frac{y - L_y / 4}{d} \right) + \tanh \left( \frac{y - 3L_y / 4}{d} \right) \right] + B_\beta, \label{eq:By}
\end{eqnarray}
where $B_\alpha = \left( \frac{B_1 - B_2}{2} \right)$, $B_\beta = \left( \frac{B_1 + B_2}{2} \right) $, $B_1 = 0.16 B_0$, and $B_2 = 0.0$. The density is set with $\rho_1 = \rho_0$ and $\rho_2 = 0.73 \rho_0$ in normalized units. The initial temperature profile is set such that the total (magnetic plus thermal) pressure is balanced across the shear layer, where $T_1 = 0.02 T_0$ and $T_2 = 0.055 T_0$. Finally, the background shear is initially perturbed by adding a small compressive velocity field in the $Y$-direction in the form 
\begin{eqnarray}
d u_y = \delta u_0 \left[ e^{\left( \frac{y - L_y/4}{4d} \right)^2} - e^{- \left( \frac{y - 3L_y/4}{4d} \right)^2} \right] \text{ran}(x) \nonumber 
\end{eqnarray}
where $\delta u_0 = 0.008 (U_2 - U_1)$ (i.e., less than $10\%$ of the shear flow magnitude) and $\text{ran}(x)$ represents a random number generator in the range [-0.5, 0.5] at each grid value.  

Fig.~\ref{fig:KHsim} shows a snapshot of the simulation at $\sim 13$ large-eddy turnover time for one of the shear layers (the two shear layer develop similar KHI structures). The color represents the flow vorticity ($\omega$) in the out-of-plane direction. The KH waves are seen to develop in the simulation. They quickly reach the non-linear stage where rolled-up KH vortices clearly form (from $\sim 6$ large-eddy turnover times onwards) with visible vortex merging. This confirms that the solar wind observations by SolO are consistent with the KHI growth. In the lower part of the simulation (Side 1), there are features seen as stripes in vorticity. These features are shocks that are produced by the supersonic flow on Side 1 (Mach number $\sim 3$) as the speed difference between the recirculating vortex and the nearby passing flow exceeds the sound speed \citep[e.g.,][]{1987flme.book.....L}. Future work ought to determine whether such features are sometimes observed in spacecraft data. For the present work, the key point would remain that the KHI does develop.

\subsection{Boundary layer analysis} \label{subsec:boundary-analysis}

   \begin{figure}[ht]
   \centering
   \includegraphics[width=3.5in]{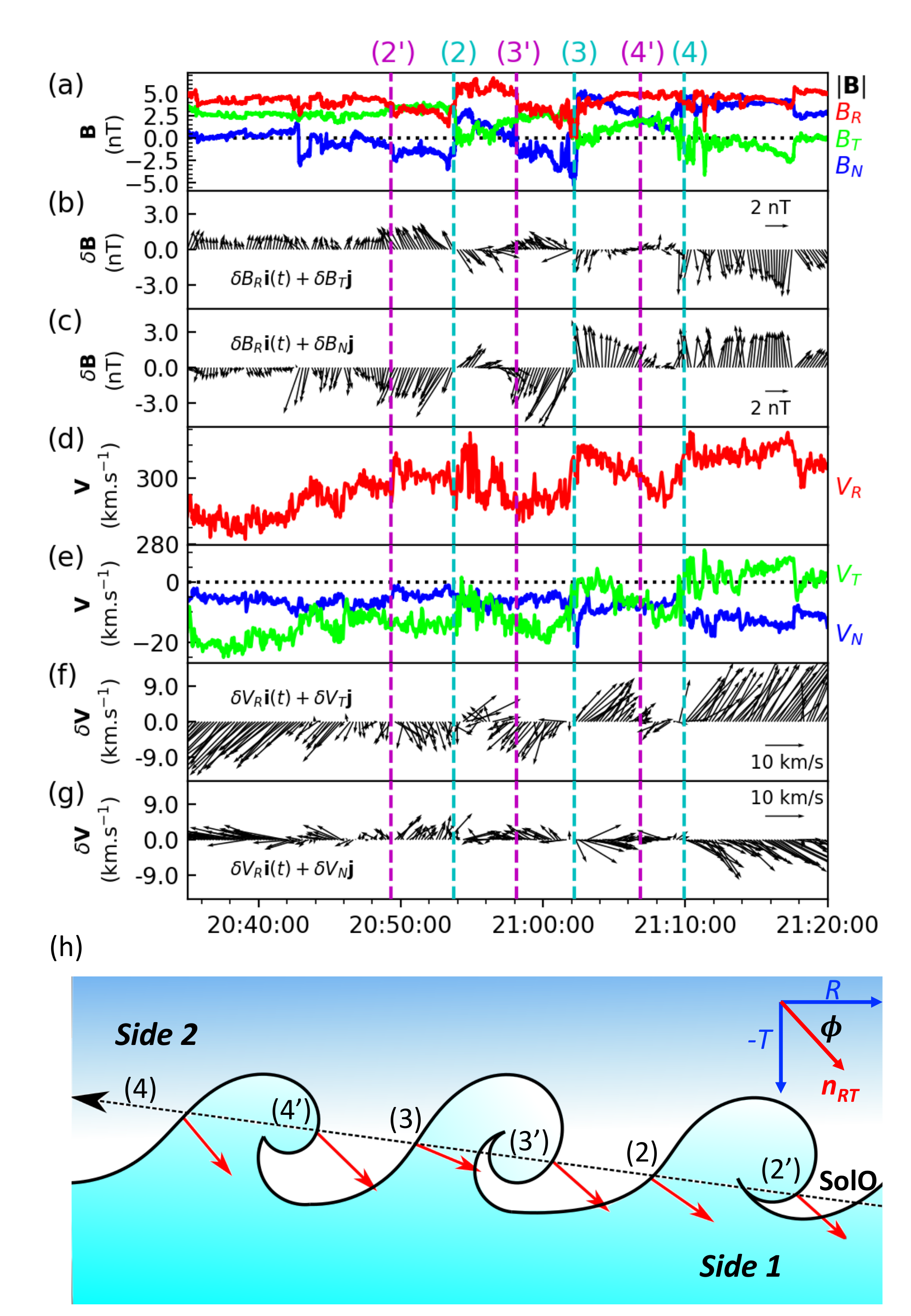}
      \caption{SolO observations on July 23, 2020 with marked outbound and inbound crossings (i.e., wave edges) of the in-situ KH waves and a schematic sketch of the event. (a) Magnetic fields. (b, c) Magnetic field fluctuations from the averages in the $R-T$ and $R-N$ planes, respectively. (d) Ion bulk velocity $V_R$ component. (e) Ion bulk velocity $V_T$ and $V_N$ components. (f, g) Velocity field perturbations in the $R-T$ and $R-N$ planes, respectively. Cyan dashed lines mark times of the outbound crossings, which correspond to KH trailing edges. Magenta dashed lines mark times of the inbound crossings, which correspond to KH leading edges. Normal directions of the inbound and outbound crossings are listed in Table~\ref{table:normal}. (h) Schematic sketch of the KH waves based on the normal angle (red arrows) at the wave edges.
              }
         \label{fig:edges}
   \end{figure}

To understand local configurations of the KH waves, we characterize orientations of the observed wave edges in the RTN coordinates. In Fig.~\ref{fig:KH}, the magnetic rotations are clearly defined for wave edges (2), (3), and (4). Fig.~\ref{fig:edges} shows $\mathbf{B}$ and $\mathbf{V}$ and their fluctuations from the average values between 20:35 and 21:20 UT. Wave edges (2) - (3), defined by clear rotations of $\mathbf{B}$ in Fig.~\ref{fig:edges}a and $\mathbf{V}$ in Figs.~\ref{fig:edges}d and \ref{fig:edges}e, are marked by cyan vertical dashed lines. We define these sharp rotations as the ``outbound crossings", which correspond to crossings of KH wave trailing edges that separate adjacent vortices. A KH trailing edge is typically thin, seen as a sharp transition, as a consequence of the compression between two waveforms or vortices. Smaller magnetic field rotations are also visible between these outbound crossings. We mark these smaller rotations as (2'), (3') , and (4') with magenta vertical dashed lines in Fig.~\ref{fig:edges}. These edges show reverse transitions compared to the outbound crossings in $B_R$ (red) and $B_N$ (blue) components. We define these smaller transitions as ``inbound crossings'',  which correspond to crossings of KH wave leading edges. A KH leading edge is typically less sharply defined in in-situ data as it is in the vicinity of KH vortices that are regions of plasma mixing, so the transitions between two regions of the shear layer are less clear \citep[e.g.,][]{1993GeoRL..20.2699C, Fairfield2000a}.

To analyze the orientations of the wave edges, we calculate the boundary normals of the inbound and outbound pairs marked in Fig.~\ref{fig:edges}. The normal of a discontinuity (i.e., current sheet) can be obtained from the cross-product of magnetic fields on either side of the discontinuity, i.e., $\mathbf{n} = \pm (\langle \mathbf{B}_1 \rangle \times \langle \mathbf{B}_2 \rangle) / |\langle \mathbf{B}_1 \rangle \times \langle \mathbf{B}_2 \rangle |$, where $\langle \mathbf{B}_1 \rangle$ and $\langle \mathbf{B}_2 \rangle$ are time averages of asymptotic magnetic fields before and after the current sheet interval, respectively. The obtained normal direction has a sign ambiguity ($\pm$); we assign a direction outward from the Sun, i.e., by forcing the radial component of $\mathbf{n}$ to be positive. The time-averaged $\langle \mathbf{B}_i \rangle$, where $i = 1,2$, are defined as $10$-s averages of the magnetic fields. We obtain the normal orientations ($\mathbf{n} = [n_R, n_T, n_N]$) of the marked inbound and outbound crossings in Fig.~\ref{fig:edges}. In addition, we define an angle $\phi = \arctan(-n_T/n_R) $ to be the angle of the current sheet normal from the $R$ direction in the $RTN$ coordinate system. The angle $\phi$ is in range $[-90^o, 90^o]$ where $\phi = 0^o$ is parallel to the Sun-SolO line and $\phi = -90^o$ and $90^0$ are perpendicular to the Sun-SolO line in the $T$ and $-T$ directions, respectively.

\begin{table}
\caption{Normal directions and orientations of the inbound (leading edge) and outbound (trailing edge) crossings with their normal directions ($\mathbf{n}$) and associated angle $\phi = \arctan(- n_T / n_R)$.}              
\label{table:normal}      
\centering                                      
\begin{tabular}{l c c c}          
\hline\hline                        
Wave edges & Times (UT) & Normal direction  & Angle $\phi$ \\    
\hline                                   
    Inbound (2')         &        20:49:20         &   [0.64, -0.57, -0.51]         &    $41.7^o$ \\  
    Outbound (2)       &        20:53:43         &       [0.78, -0.56, -0.28]      &    $35.7^o$   \\  
\hline
    Inbound (3')         &        20:58:08         &  [0.71, -0.63, -0.31]           &    $41.5^o$ \\  
    Outbound (3)       &        21:02:11         &       [0.91, -0.36, -0.21]        &   $21.6^o$   \\  
\hline       
     Inbound (4')         &        21:06:50         &   [0.70, -0.68, 0.2]            &     $44.2^o$ \\  
    Outbound (4)       &        21:09:56         &    [0.58, -0.67, 0.45]          &    $49.1^o$  \\   
\hline\hline
\end{tabular}
\end{table}

Table~\ref{table:normal} notes times of the inbound and outbound pairs, their normal directions ($\mathbf{n}$), and the normal angle $\phi$. The normal directions for all inbound and outbound crossings are in between the $R$ and $-T$ directions; these translate to the normal angles between $0^o$ and $90^o$. The normal angles ($\phi$) of an inbound-outbound pair were used to characterize in-situ waves at the Earth's magnetopause by \citep{Plaschke2016a}. By considering the angles of an inbound and outbound pair, we can determine whether the in-situ waveform is consistent with a sinusoidal wave, a KH wave, or another type of wave. Of particular relevance is that, when the waveform is a KH vortex (i.e., a rolled-up KH wave), the angles of the inbound and outbound pair should be in the same range (cf. Fig.~1 of \citet{Plaschke2016a}). The observed angles of all the inbound-outbound pairs in Table~\ref{table:normal} are well consistent with the steepening of KH vortices. This result supports the argument that the KHI is developed at the shear layer. We sketch the wave trains based on the normal angles with a hypothetical relative SolO trajectory in Fig.~\ref{fig:edges}h. Note that, in reality, the spacecraft is relatively static as the structures flow past with the solar wind.

To further see the vortex-type variations in $\mathbf{B}$ and $\mathbf{V}$, we define their fluctuation vectors from the averages ($\delta \mathbf{B}$ and $\delta \mathbf{V}$) between 20:35 and 21:20 UT. Figs.~\ref{fig:edges}b and ~\ref{fig:edges}c show $\delta \mathbf{B}$ in the $R-T$ and $R-N$ planes, respectively. Around the inbound crossing (3'), $\delta \mathbf{B}$ in the $R-N$ plane rotates smoothly in a clockwise sense. This pattern is also observed at other inbound crossings, although it is less pronounced. Figs.~\ref{fig:edges}f and ~\ref{fig:edges}g show $\delta \mathbf{V}$ in the $R-T$ and $R-N$ planes, respectively. A clockwise pattern is also visible between (2) and (3') in Fig.~\ref{fig:edges}f. This pattern of rotation seen in both $\delta \mathbf{B}$ and $\delta \mathbf{V}$ is consistent with a vortical structure (i.e., a rolled-up KH vortex). The pattern is not as obvious at other inbound crossings; this may indicate that SolO does not pass centrally through the KH vortices in these cases.

\subsection{Magnetic spectra} \label{subsec:turbulence}

\begin{figure}
   \centering
     \includegraphics[width=3.5in]{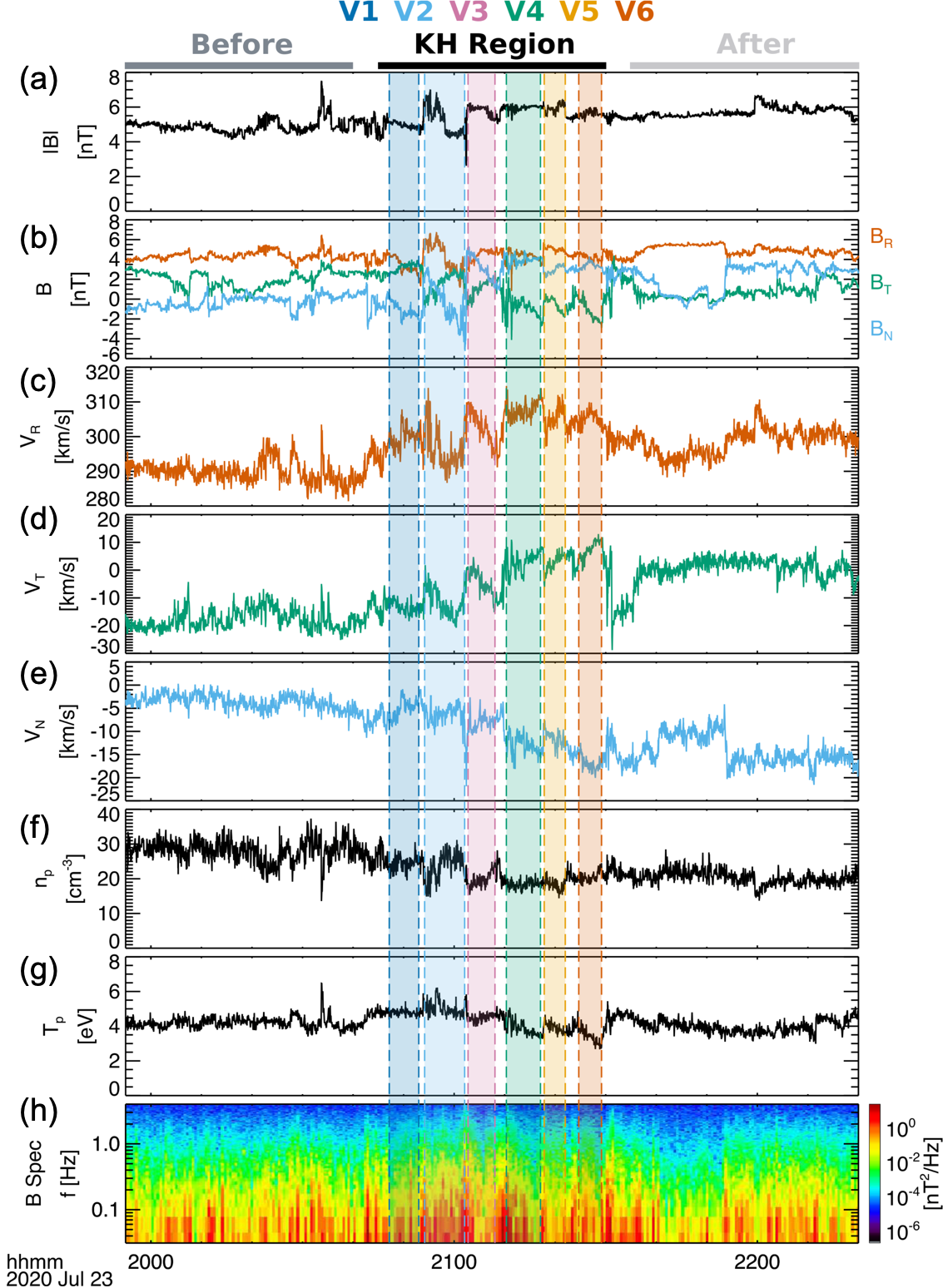}
      \caption{Overview of the intervals for magnetic spectrum analysis. The KH region is marked in the middle (see top) together with the intervals before (top left) and after (top right). The KH vortices V1 to V6, marked between the compressed current sheets, are shaded in colors. (a) Total magnetic field. (b) Magnetic field in the RTN system. (c, d, e) Velocity field $R$, $T$, and $N$ components, respectively. (f) Ion number density. (g) Ion temperature. (h) Magnetic spectrum. 
              }
         \label{fig:KH-turbulence-overview}
   \end{figure}

We now examine turbulence properties of the KH event. Fig.~\ref{fig:KH-turbulence-overview} shows the KHI interval with KH sub-regions V1 to V6 highlighted with colors (middle) together with time periods before and after the KH interval (see top). Fig.~\ref{fig:KH-turbulence-overview}h shows a spectrogram of magnetic spectrum. The magnetic spectrum shows enhancement within the KH region compared to before or after the interval. The enhancement is visually strongest in V2 compared with other vortices. This V2 is the same interval as between (2) and (3') in Fig.~\ref{fig:edges} where we see a clear clockwise rotation in the magnetic field and velocity field perturbations consistent with a rolled-up KH vortex. Thus, the strong enhancement in magnetic wave power provides evidence of enhanced activity in this vortex as plausibly facilitated by the development of a nonlinear KH vortex. The enhancement of the magnetic spectrum in V4 is also strong but less than for V2.

      \begin{figure}
   \centering
   \includegraphics[width=3.5in]{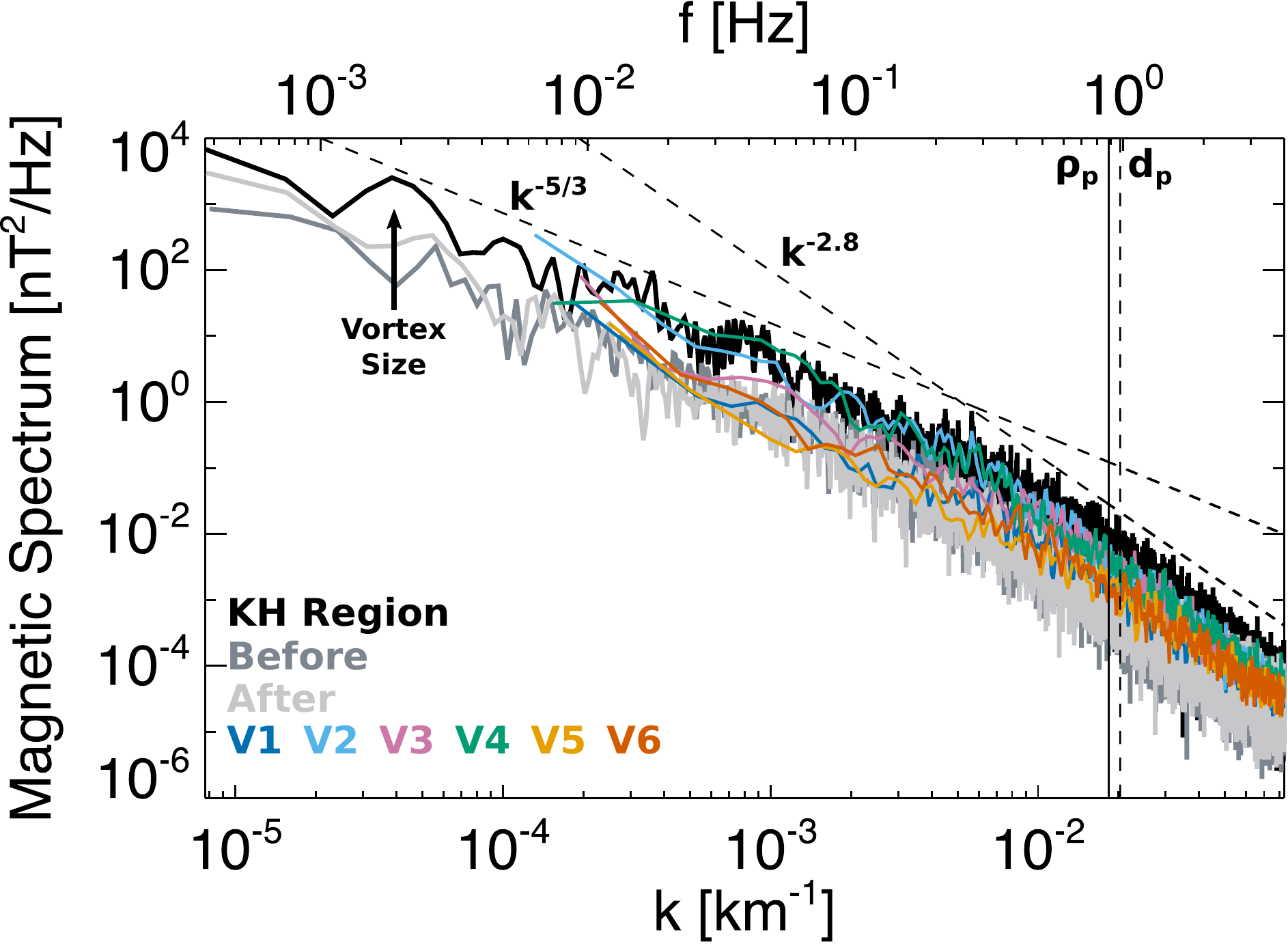}
      \caption{Magnetic spectra of all regions marked in Fig.~\ref{fig:KH-turbulence-overview}. The Kolmogorov power law $k^{-5/3}$ and the dissipation range scalings $k^{-2.8}$ are plotted for reference as straight black dashed lines. The vortex size is noted by an arrow (top left). The ion gyroradius scale ($\rho_p$) and the ion inertial length ($d_p$) are marked by vertical black solid and dashed lines, respectively. 
              }
         \label{fig:KH-spectra}
   \end{figure}

To quantitatively assess the magnetic field fluctuations in the full KHI region and in each of the six sub-regions V1-V6 marked in Fig.~\ref{fig:KH-turbulence-overview}, we computed magnetic field spectra, which are shown in Fig.~\ref{fig:KH-spectra}. Additionally, the KH-stable regions immediately before and after the KHI event are shown for reference. The vortex size is marked by an arrow (top left). The spectrum for the full KH region is shown in black while the spectra of the intervals before and after are shown in grey. As references, power law scalings of $k^{-5/3}$ \citep{Kolmogorov1941} at MHD scales and $k^{-2.8}$ at ion scales \citep[i.e.,][]{2009PhRvL.103p5003A} are plotted as black dashed lines. The scale of a thermal ion gyroradius ($\rho_p$) and an ion inertial length ($d_p$) based on the average properties in the KH region are marked with vertical solid and dashed black lines, respectively. The spectrum of the entire KH interval has more power than the intervals before and after, which both have similar low-intensity spectra. This suggests that the KHI is exciting additional fluctuations. The magnetic spectrum of the KH region approximately follows both power laws with a spectral breakpoint at $f \sim 0.2$ Hz. The spectrum essentially follows the power law $k^{-2.8}$ for scales smaller than the ion gyroradius. This indicates that the magnetic spectrum of the KHI interval is consistent with a classic turbulence cascade down to the kinetic scales \citep[e.g.,][]{2014ApJ...793L..15B, 2017MNRAS.472.1052B}. This result provides evidence of shear-driven turbulence as driven by the local KHI.

Fig.~\ref{fig:KH-spectra} also shows magnetic spectra for individual vortices V1 - V6. Note that the compressed current sheets are excluded for the analyses of these vortices (this is why the colored regions are not exactly contiguous on Fig.~\ref{fig:KH-turbulence-overview}). The powers of the magnetic spectra of all vortices are weaker than that of the entire KH region, indicating that the current sheets are key regions for enhancing the power spectrum. V2 (blue) and V4 (green) appear to have higher powers compared to other vortices and almost reach the power of the entire KH region (black). The enhanced power in these vortices may be related to the excitation of turbulent fluctuations through secondary instabilities and the nonlinear evolution of the KHI. The difference in the power spectrum between the different vortices may indicate that SolO was crossing different parts of KH vortices while crossing through the shear layer from Side 1 to Side 2. It is likely that SolO was passing through the center of a rolled-up vortex in V2, for reasons noted earlier. The lower powers of V1, V5, and V6 may indicate that SolO was skimming through the trough or crest parts of KH vortices. These results are in agreement with the assumed spacecraft trajectory through the KHI structure using the boundary layer analysis in Section~\ref{subsec:boundary-analysis}.


\section{Discussion} \label{sec:discussion}

We have reported observations of the KHI within a shear layer embedded in the slow solar wind close to an HCS using Solar Orbiter. The event is observed in the inner heliosphere at a distance of $\sim 0.69$ AU. Despite several theoretical postulations \citep[e.g.,][]{Parker1963, Sturrock1966, Miura1982, Korzhov1984, Neugebauer1986, Hollweg1987} and spacecraft missions in the inner heliosphere, direct evidences for the KHI were not reported in past in-situ observations of the solar wind. We discuss why this event may be favorable for a KHI detection as well as implications of the KHI in the solar wind as follows.

\subsection{KHI criterion in the solar wind} \label{diss:KHI-onset}

First, we consider solar wind conditions that are favorable for the KHI. Based on the KHI onset condition (Eq.~\ref{eq:onset}), the shear layer can more easily become unstable to the KHI when $B$ is low and $n$ is high because it requires a velocity jump across the shear layer greater than the local Alfvén speed ($\Delta V > V_A$). Near the Sun, $B$ and $V_A$ are typically large. This inhibitory effect on the KHI criterion should be stronger near the Sun. Nevertheless, there may be several situations where the local conditions allow the KHI. For example, remote sensing by \citet{2018ApJ...862...18D} show a particularly strong shear values of a $\Delta V_R \sim 200$ km s$^{-1}$ across streamer structures in the young solar wind. \citet{Ruffolo2020} also propose the KHI development near the Alfvén critical zone, at $R < 0.17$ AU. The event we analyze here was observed near the HCS with many coherent structures and shears. Besides, this event is found in the slow wind, which is generally dense, making the conditions to meet the KHI criterion easier.

\begin{figure}
   \centering
     \includegraphics[width=3.5in]{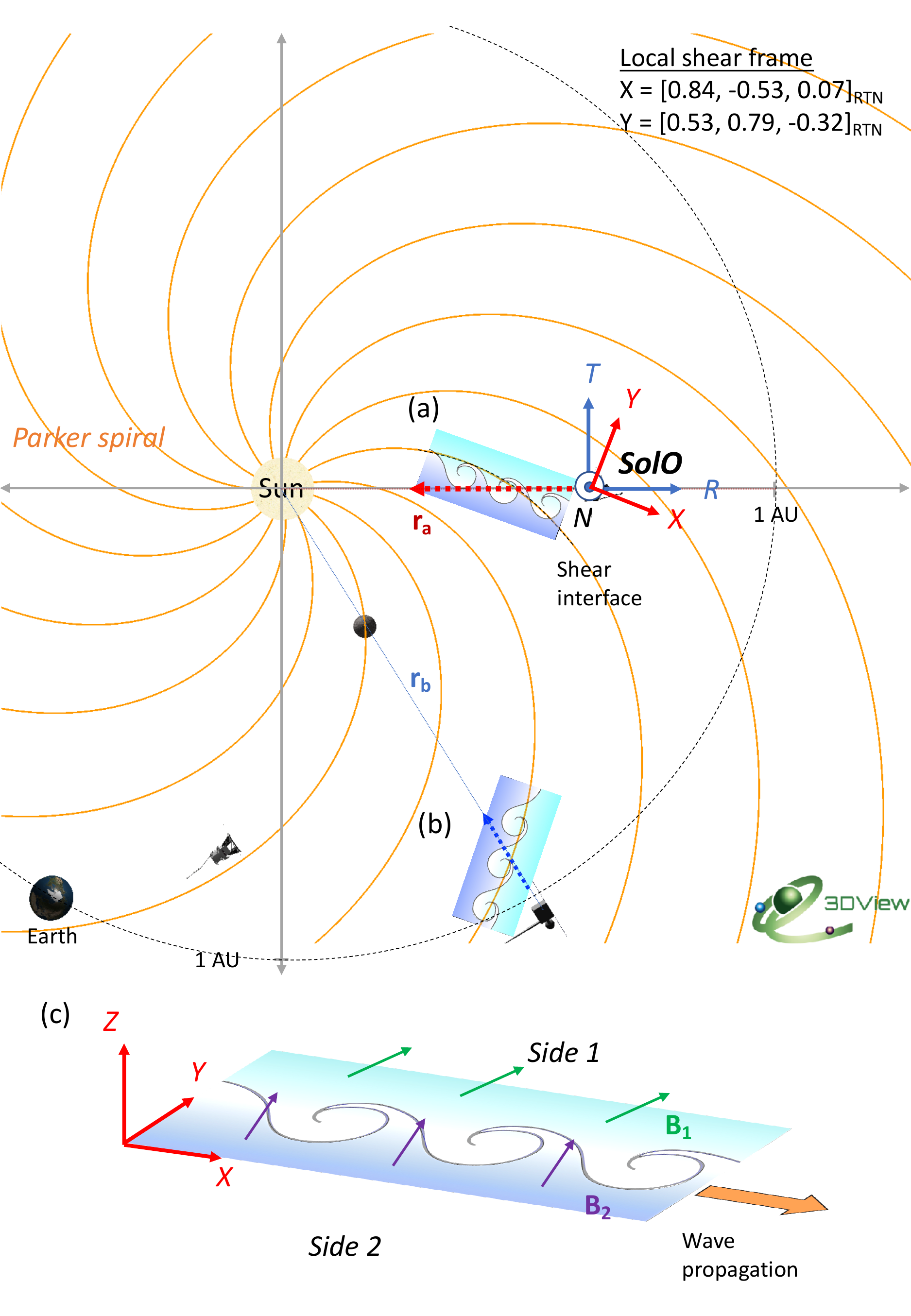}
      \caption{(top) Schematic illustrations of scenarios for which the KHI develops (a) near SolO position and (b) at $1$ AU, shown in the ecliptic plane with the Parker spirals shown with orange solid lines (similar to Fig.~\ref{fig:SOpos}). At (a), we project the derived $X-Y$ plane onto the $R-T$ plan as described in the text. The main flow is radially outward from the Sun as marked with vectors $\vec{r_a}$ and $\vec{r_b}$. As the Parker spiral becomes more bent at (b), the spacecraft relative spacecraft trajectory is more perpendicular to the shear interface, making the likelihood of crossing several vortices much lower than in the inner heliosphere at (a). (bottom, c) A simplistic 3-D view of the KHI in (a) with the out-of-plane ($Z$) magnetic field shown with arrows on Sides 1 and 2, with corresponding values in Fig.~\ref{fig:MVA-box}e. 
              }
         \label{fig:sketch}
   \end{figure}

Second, we consider the magnetic and velocity field configurations across shear layers. The KHI is suppressed when $\mathbf{B}$ is strong in the direction of $\Delta \mathbf{V}$ due to the stabilization by magnetic tension in the direction of the shear flow. In this interplanetary medium, we typically expect a shear interface along the Parker spiral direction (see Fig.~\ref{fig:sketch}) so $\mathbf{B}$ may usually be aligned with $\Delta \mathbf{V}$. However, near the Sun, there may be velocity shear due to the solar-wind corotation with the Sun. Several studies have shown that the KHI may occur in various situations, i.e., at the edge of a CME \citep[e.g.,][]{Foullon2013, Mostl2013} and at the interfaces between CME and sheath and between sheath and solar wind \citep[e.g.,][]{Paez2017}.

Third, Eq.~\ref{eq:onset} is derived by assuming an ideal MHD plasma with an infinitely thin shear layer. In reality, non-ideal MHD effects such as the compressibility can stabilize the KHI \citep[e.g.,][]{Sen1964}; for example, the KHI only grows for a limited range of the velocity jump across a shear flow for a 1-D TD in homogeneous plasmas and magnetic fields \citep[e.g.,][]{Talwar1964, Pu1983}. The solar wind is indeed compressible and thus we expect some stabilizing effects. In addition, shear layers have finite thicknesses. A finite thickness of the shear layer can also stabilize the KH mode for small wavelength perturbations (i.e., for large wave number $k$). A combination of the compressibility and the finite thickness can stabilize the KHI such that only certain modes of $k \Delta L$, where $\Delta L$ is the shear layer thickness, are KHI unstable \citep{1982JGR....87.7431M}. Although these two factors can impact the shear-layer stability, we do not expect their effects to be large, nor to be specifically dependent on distance from the Sun.

To summarize, there are factors that can impact the KHI development in the solar wind. The magnitude of $B$ and $V_A$ depend on distance from the Sun. As $V_A$ is higher closer to the Sun, the KHI criterion should be more difficult to satisfy, except where the shear $\Delta V$ is particularly strong. Often $\mathbf{B}$ may be parallel to the velocity shear, tending to inhibit the KHI, except in some circumstances, e.g., when there is a CME that changes the local conditions. Compressibility of the solar wind and a finite thickness of the shear region can help stabilize the KHI. 

Since the observed conditions during our event are not particularly unusual for dense solar wind near the HCS, KHI development should not be rare. We now consider arguments related to KHI timescale and configuration as follows.

\subsection{KHI timescale}

A first fact to consider now is that when the KHI develops at a shear layer, it quickly reaches the non-linear stage (i.e., the rolled-up stage). The periodic features and vortical structures then get rapidly destroyed as the plasmas from either side of the shear layer mix and vortices coalesce. At such late stage, they would be indistinguishable from other solar wind types, albeit likely associated with higher levels of fluctuations as we have actually found in Section~\ref{subsec:turbulence} for this event. The timescale for the decay of a KH vortex is on the order of one to a few eddy turnover time. The eddy turnover time of a KH vortex scales as a fraction of the KH wavelength ($\lambda_{KH}$) divided by the KH wave phase speed ($V_{ph}$). Assuming that the KH vortex size is about $\lambda_{KH}/3$ to $\lambda_{KH}/2$ ($22,000$ to $33,000$ km) for the fastest growing mode \citep[i.e.,][]{Walker1981, Miura1982}, with the $V_{ph} = 152$ km s$^{-1}$ (see Section~\ref{subsec:linear-theory}), we obtain a KH vortex turnover time of $2.4$ to $3.6$ minutes. Since the conditions of our event are not unusual for dense solar wind near the HCS, this short timescale for turnover time (before the vortex decay) may contribute to the rarity of KHI detection.

It is also possible that KH waves were detected by past missions but their signatures were not resolved. Several periodic oscillations in magnetic field strength were observed by \cite{Burlaga1968} using the Pioneer-6 spacecraft, launched in 1965, at $\sim 0.8$ AU. One of the cases considered was found to have sinusoidal $|\mathbf{B}|$ oscillations with a period of $\sim 5$ minutes, embedded in a velocity shear layer. Although no other fluctuations were seen in the data, it was suggested that these waves were generated by a KHI.

\subsection{KHI configuration}
   
When TDs form in the solar wind, they are typically at the boundary between solar wind flux tubes that move with relative velocity. Since the solar wind flux tubes bend following Parker spiral arcs to first order, we expect the normal of shear interfaces to be generally perpendicular to the Parker's spiral. Fig.~\ref{fig:sketch} shows a schematic illustration of expected KHI configurations with respect to the solar wind Parker spiral in the ecliptic plane, produced using a constant velocity of $300$ km s$^{-1}$ close to the observed speed. We illustrate two scenarios of the KHI configuration (a) in the interplanetary medium between $0.3$ and $0.7$ AU and (b) near $1$ AU. We describe a scenario for which the KHI develops relative to the Parker spiral based on our data as follows.

For our event at $\sim 0.7$ AU depicted at (a), we found that the velocity jumps for a displacement is in the direction $Y = [0.53, 0.79, -0.32]_{RTN}$ (the maximum variance direction of $\mathbf{V}$). This direction is indeed perpendicular to the Parker spiral, which is rather oriented along what we call the $X$ direction, as shown at (a) (see frame definition in upper right part of the Fig.~\ref{fig:sketch}). The KH wave amplitude spreads in the Y direction as it is perpendicular to the shear interface. The wave propagation direction is $X = [0.84, -0.53, 0.07]_{RTN}$ (the intermediate variance direction of $\mathbf{V}$). We expect that the KHI configuration in the solar wind will often be similar to this geometry. As explained next, such a KHI configuration relative to a more slowly moving spacecraft in the inner heliosphere impacts the likelihood of in-situ observations of the KHI.

We now consider two scenarios of KHI development (a) near SolO position at $\sim0.7$ AU and (b) near 1 AU in Fig.~\ref{fig:sketch}. As the main solar wind flow is in the radial direction, a static spacecraft at (a) would sample several vortices within the shear layer only if it has an appropriate relative trajectory through the layer. Note that SolO was relatively static (with a speed of $11$ km s$^{-1}$) compared to the KH wave phase speed ($152$ km s$^{-1}$). At a further distance from the Sun, i.e., near $1$ AU as shown in (b), the Parker's spiral becomes more bent. The shear interface thus becomes more tilted relative to the outward solar wind flow as compared to the geometry observed at locations in the inner heliosphere (as show in Fig.~\ref{fig:sketch}a). Consequently, a spacecraft located in the inner heliosphere has a relative trajectory that makes it more likely to sample several vortices than a spacecraft that is farther out in the heliosphere, and which will make a more perpendicular cut across the shear layer, as depicted in Fig.~\ref{fig:sketch}b. This may be a prime explanation as to why despite the many spacecraft near $1$ AU, there were no reports of in-situ KHI detection.

Note that the above considerations are based on a simplistic 2-D view. In reality, 3-D effects such as out-of-ecliptic magnetic and velocity fields can significantly impact the KHI configuration and development with respect to the Parker spiral. For example, we show magnetic fields perpendicular to the shear plane in Fig.~\ref{fig:sketch}c and these are in fact required to avoid the alignment of the velocity and magnetic shear, which otherwise has a stabilizing effect on the KHI. We acknowledge that our considerations are to first order only; future work should address this in more detail.

\subsection{Implications of the KHI in the solar wind}

The KHI is expected to play important roles, such as allowing for plasma mixing, generating turbulence, or producing Alfvénic fluctuations in the solar wind as mediated by KH vortex dynamics. During the present event, SolO observed an ion jet consistent with magnetic reconnection (see Section~\ref{subsec:recon}). An interesting question is whether this reconnection is produced due to dynamics of a KH vortex. For vortex-induced reconnection (VIR), we expect reconnection to be produced at a thin current sheet in between two vortices.  At the Earth's magnetopause, VIR jets were found to orient themselves along KH trailing edges \citep[e.g.,][]{Eriksson2016a} as the vortex evolves and further enhances the magnetic shear. In our case, we found that the jet is in the out-of-ecliptic direction while the KH trailing edges (see Table.~\ref{table:normal}) and the shear layer are in the $R, T$ directions. Thus, it is unclear whether the observed jet is a VIR. We think that the $N-$ directed jet is rather a consequence of the inclination of the local current sheet. Nevertheless, the KH vortex may further increase the magnetic shear at the KH edge and make the current sheet thin enough to trigger magnetic reconnection.

We now discuss the magnetic and velocity field fluctuations. We found that the KHI enhances fluctuations compared to outside the interval and the magnetic spectrum of the KHI region approximately follows the power law scalings of $k^{-5/3}$ and $k^{-2.8}$ at inertial and kinetic scales, respectively (Section~\ref{subsec:turbulence}). These enhanced fluctuations are consistent with turbulence generation by the KHI at the magnetopause \citep[e.g.,][]{2016JGRA..12111021S, 2017NatCo...8.1582N}. Therefore, the magnetic spectrum is consistent with a classical turbulence cascade down to the kinetic scales. These observations are consistent with an enhancement of turbulence in the solar wind as driven by the local KHI \citep[e.g.,][]{1989GMS....54..113G}. We also note that current sheets are key structures that contribute to the power spectrum, as power spectra of only vortex regions are lower than the overall spectrum. In addition, several of the vortices have enhanced power within them, which may be due to secondary instabilities, perhaps supporting the idea that the KHI helps to drive some fraction of the turbulent fluctuations in the solar wind.

One important implication of KHI in the solar wind is that it can contribute to the evolution of the magnetic and velocity fluctuations. Near the Sun, the KHI is believed to be a mechanism that leads to shear-driven turbulence at the Alfvén critical zone where $V = V_A$ and in the vicinity of the $\beta = 1$ surface \citep{DeForest2016, Chhiber2018}, leading to more isotropic solar-wind streams. Furthermore, the dynamical evolution invoked by shear-driven instabilities such as the KHI is found to be able to account for features observed by PSP including magnetic ``switchbacks'' near perihelia  \citep{Ruffolo2020}. This topic needs to be investigated further but it is beyond the scope of the present study.

\section{Conclusions} \label{sec:conclusions}

We report observations of the KHI with SolO on July 23, 2020 at $0.69$ AU, during the cruise phase. The KH waves are observed within the velocity shear layer with periodic fluctuations in several parameters in the slow solar wind near an HCS. Several KH waveforms are observed with a period of $\sim 7$ minutes but only a few vortices are clearly noticed. One of them is found to have magnetic and velocity field perturbations consistent with a rolled-up KH vortex in the non-linear stage. We test the observed conditions on either side of the shear layer with linear theory and find that the shear layer is indeed KHI-unstable. The maximum variance direction of the shear flow, associated to the KH wave amplitude direction, is found in the direction approximately perpendicular to the Parker spiral. The intermediate variance direction of the shear flow, associated with the net flow and thus the direction of the KH wave propagation, is along the Parker spiral. Using linear theory, the wave phase speed is estimated to be $152$ km s$^{-1}$. The KH wavelength is approximately $66,400$ km or $0.1$ solar radii. We also comfirm the local KHI development by exploiting a 2-D MHD simulation with the empirical values.

Additionally, we report the observation of an ion jet consistent with magnetic reconnection at one of the outbound (trailing) edges, likely as a result of current sheet compression in between two KH vortices. The ion jet has $\Delta V = 11$ km s$^{-1}$ along the maximum variance direction, consistent with magnetic reconnection with a sub-Alfvénic jet. Nevertheless, we found other signatures consistent with magnetic reconnection, namely a drop in magnetic field strength, an ion number density enhancement \citep{Gosling2005}, and plasma heating \citep{2014GeoRL..41.7002P}. It is unclear whether this jet is produced due to KH vortex-induced reconnection \citep[e.g.,][]{Nakamura2006} or the local inclination of the magnetic field, as the jet direction does not correspond to the KH trailing edge orientation \citep{Eriksson2016a}. 

We also report the enhancement of the magnetic and velocity field fluctuations within the KHI interval compared to intervals before and after. The power of the magnetic spectrum of the entire KHI interval approximately follows the power law scalings of $k^{-5.3}$ and $k^{-2.8}$ in the inertial and kinetic ranges, respectively, consistent with the turbulent cascade in the solar wind. This provides evidence for the local enhancement of turbulence as driven by the KHI.  Moreover, we find that current sheets within the KHI interval are key structures that enhance the power, as the magnetic spectra of individual KH vortices (excluding compressed current sheet intervals) generally have less power.

As our reported event here is an unambiguous in-situ observation of the KHI in the solar wind, we discuss possible reasons why the KHI was not reported in past in-situ observations. First, the KHI onset criterion requires a velocity jump across the shear layer that is larger than the local Alfvén speed ($\Delta V > V_A$) and weak magnetic field in the direction of the shear flow (i.e., low $\mathbf{B} \cdot \mathbf{V}$). Second, the KHI is estimated to quickly reaches the nonlinear stage where KH vortices roll up and merge. This timescale should be on the order of a few eddy turnover. The observed conditions are typical in the solar wind, and we estimate the timescale of the KHI to be on the order of minutes. In other words, when KHI develops in the solar wind, it evolves rapidly and is thus rather ephemeral. Third, the configuration in which the KHI develops may not be optimal for detection by a relatively slow-moving spacecraft near $1$ AU owing to the geometry of the Parker spiral. The KHI develops at discontinuities formed in between solar wind flux tubes that move with a relative velocity. In this picture, the KH wave amplitude should spread in a direction perpendicular to the Parker spiral, albeit often with significant north-south components as in our case. Owing to its more skimming trajectory, a spacecraft in the inner heliosphere such as SolO may more easily pick up several waveforms in the shear layer. Further from the Sun, in contrast, the Parker spiral becomes more bent. A spacecraft near $1$ AU is less likely to sample several vortices along its trajectory which has a larger angle to the discontinuity, making it difficult to recognize KH waves. We think these are main factors that contribute to the rarity of  in-situ KHI detection in the solar wind.

This event provides evidence for the existence of the KHI in the solar wind. It sheds new light on solar wind shear processes in the interplanetary medium with direct applications to shear-driven turbulence mediated by the KHI, likely contributing to the solar wind fluctuations observed at $1$ AU \citep[e.g.,][]{Ruffolo2020}. As the Alfvén speed decreases away from the Sun, the KH growth rate becomes higher \citep[e.g.,][]{Neugebauer1986} and thus the KHI may be more common. Due to the short timescale of the linear KHI, there may be more chances to detect nonlinear KHI or its remnants. Recently, techniques for detecting kinetic features of the KHI during the nonlinear and turbulent stage of the KHI were proposed \citep{2021arXiv210204117S}. Further studies would be needed to study secondary processes induced by the KHI such as vortex-induced reconnection and other kinetic mechanisms, as the KHI is rich with magnetic and plasma structures as is well known for the case of the magnetopause.

\begin{acknowledgements}
Work at IRAP was supported by the Centre National de la Recherche Scientifique (CNRS, France), the Centre National d’Etudes Spatiales (CNES, France), and the Université Paul Sabatier (UPS). Solar Orbiter data are publicly available at \url{http://soar.esac.esa.int/soar/}. We acknowledge science teams of the Connectivity Tool (\url{http://connect-tool.irap.omp.eu/}) and 3DView (\url{http://3dview.irap.omp.eu/}) at IRAP and CNES. Y. Y. is supported by grant No. 11902138 from the National Natural Science Foundation of China. The computing resources were provided by the Center for Computational Science and Engineering of Southern University of Science and Technology. D.R. is supported by grant RTA6280002 from Thailand Science Research and Innovation. J.E.S is supported by the Royal Society University Research Fellowship URF$\backslash$R1$\backslash$201286. C.J.O. is funded under STFC grant number ST/5000240/1. Solar Orbiter Solar Wind Analyser (SWA) data are derived from scientific sensors which have been designed and created, and are operated under funding provided in numerous contracts from the UK Space Agency (UKSA), the UK Science and Technology Facilities Council (STFC), the Agenzia Spaziale Italiana (ASI), the CNES, the CNRS, the Czech contribution to the ESA PRODEX programme and NASA. Solar Orbiter SWA work at UCL/MSSL is currently funded under STFC grants ST/T001356/1 and ST/S000240/1.

\end{acknowledgements}

\bibliographystyle{aa} 

\end{document}